\documentclass[manuscript]{aastex}
\usepackage{graphicx}
\usepackage{lscape}
\usepackage{natbib}
\usepackage{xspace}

\newcommand{\fco}{$f_{\rm co}$\xspace}

\newcommand{\lat}{$\ell$\xspace}

\newcommand{\degre}{$^{\circ}$\xspace}

\slugcomment{submitted second revised version to Icarus (23 June 2010)}

\shorttitle{CO in Hale-Bopp } \shortauthors{Bockel\'ee-Morvan \&
Boissier}

\begin{document}


\title{No compelling evidence of distributed production of CO in comet C/1995 O1
(Hale-Bopp)  from millimeter interferometric data \\
and a reanalysis of near-IR lines}


\author{Dominique Bockel\'ee-Morvan}


\affil{LESIA, Observatoire de Paris, F-92195 Meudon, France}
 \email{dominique.bockelee@obspm.fr}

\author{J\'er\'emie Boissier}

\affil{IRAM, 300 rue de la Piscine, Domaine universitaire,
          F-38406, Saint Martin d'H\`eres, France}

\author{Nicolas Biver}

\affil{LESIA, Observatoire de Paris, F-92195 Meudon, France}

\and

\author{Jacques Crovisier}

\affil{LESIA, Observatoire de Paris, F-92195 Meudon, France}

Running head : CO in comet Hale-Bopp

Number of pages : 55

Number of Tables : 3

Number of figures : 13

Key words: comet Hale-Bopp; Comets, composition; Radio observations; Infrared observations\\

\underline{Send correspondence to:}

Dominique Bockelée-Morvan

Observatoire de Meudon

5 place Jules Janssen

F92195 Meudon, Cedex, France

Phone : 33 1 45 07 76 05

\clearpage
\begin{center}
{\bf ABSTRACT}
\end{center}
 \small{ Based on long-slit infrared spectroscopic
observations, it has been suggested that half of
   the carbon monoxide present in the atmosphere of comet C/1995 O1 (Hale-Bopp) close to perihelion was
   released by a distributed source in the coma, whose nature (dust or gas) remains unidentified. We
   re-assess the origin of CO in Hale-Bopp's coma from millimeter interferometric data and a re-analysis of the
   IR lines.

   Simultaneous observations of the CO $J$(1--0) (115 GHz)
   and $J$(2--1) (230 GHz) lines were undertaken with the IRAM Plateau de Bure interferometer in single-dish
   and interferometric modes. The diversity of angular resolutions (from 1700 to 42\,000 km diameter
   at the comet) is suitable to study the radial distribution of CO and detect the extended source
   observed in the infrared.
   We used excitation and radiative transfer models to simulate the single-dish and interferometric
   data. Various CO density distributions were considered, including 3D time-dependent hydrodynamical
   simulations which reproduce temporal variations caused by the presence of a CO rotating jet.
   The CO $J$(1--0) and $J$(2--1) observations can be consistently explained by a nuclear
   production of CO.  Composite 50:50 nuclear/extended productions with characteristic scale lengths of CO
   parent $L_p$ $>$ 1500 km are rejected.

   Based on similar radiation transfer calculations, we show that the
   CO $v$ = 1--0 ro-vibrational lines observed in comet Hale-Bopp at heliocentric distances less than 1.5 AU
   are severely optically thick. The broad extent of the CO brightness distribution in the infrared is mainly due to
   optical depth effects entering in the emitted radiation. Additional factors can be found in the complex structure of the CO
   coma, and non-ideal slit positioning caused by the anisotropy of dust IR
   emission.

   We conclude that both CO millimeter and infrared lines do not provide compelling
   evidence for a distributed source of CO in Hale-Bopp's atmosphere.
}

\noindent {\bf Keywords:} Comet Hale-Bopp -- Comets, composition
-- Radio observations -- Infrared observations

\section{Introduction}

About two dozen molecules have been detected in cometary
atmospheres in addition to a wealth of radicals, atoms and ions
formed by photodissociation, photoionization and coma chemistry
\citep{domi2004,Feld04}. It is generally agreed that most
molecules found in the coma are released from the nucleus, being
present in condensed form or as trapped gases in water ice.
However, there are pieces of evidence that distributed sources
(also called "extended sources") of gas are present in cometary
atmospheres \citep[see the reviews of][]{dbm02,cot08}. Molecules
released from the nucleus and from extended sources exhibit
different radial distributions in the coma. Molecules produced
from the thermal or photo-degradation of organic grains should
display characteristic production curves versus heliocentric
distance \citep{fray06}. Radial distributions not consistent with
nucleus production were found for H$_2$CO, CO, and OCS
\citep[e.g.,][]{meier93,disanti03,dello98}. Strong heliocentric
variations of the H$_2$CO, HNC, and CS mixing ratios were
measured, suggesting a temperature-dependent production mechanism
for these species \citep[e.g.,][]{biver99b,fray06,milam06,lis08}.
The production of radicals (CN, C$_2$) from grains is also
suggested. The study of extended sources of gases can provide
important clues to the chemical nature of refractory organic
compounds in cometary nuclei \citep{cot08}. For example,
laboratory experiments show that the decomposition of
polyoxymethylene, a polymer of formaldehyde, can be the source of
monomeric H$_2$CO in cometary atmospheres \citep{cot04,fray06}.

The first evidence for an extended CO production was obtained from
measurements of the local CO density in comet 1P/Halley along the
path of the Giotto spacecraft acquired with the neutral mass
spectrometer (NMS). \citet{ebe87} concluded that only about
one-third of the CO was released from the nucleus, the remainder
being produced by an extended source in the coma. The total CO
production rate in comet 1P/Halley was estimated to be 11\% that
of water at the time of the Giotto flyby \citep{ebe87}. An
alternative view, that Giotto flew through a jet enriched in CO,
was proposed by \citet{gree98} to interpret the NMS observations.

Constraining the radial distribution of CO in cometary atmospheres
from telescopic observations requires significant spatial
resolution and a productive comet. Long-slit spectroscopic CO
observations near 4.7 $\mu$m performed at $r_h$ $<$ 1.5 AU in
comet C/1995 O1 (Hale-Bopp) have suggested that 50 to 90 \% of the
CO in this comet was originating from a distributed source
\citep{disanti99,disanti01,bro03}. Similar observations performed
in C/1996 B2 (Hyakutake) indicate that CO was released almost
entirely ($\sim$ 75\%) from the nucleus \citep{disanti03}. CO was
abundant in these two comets with a total CO production rate
$\sim$ 20\% relative to water.

We present here the study of the radial distribution of CO in
comet Hale-Bopp from observations of the $J$(1--0) (115 GHz) and
$J$(2--1) (230 GHz) lines undertaken with the IRAM Plateau de Bure
interferometer (PdBI) \citep{bock+09}. Observations were performed
both in single-dish and interferometric modes with angular
resolutions probing coma radii from 830 to 20800 km, which
encompass the characteristic scale (5000 km) of the CO extended
source identified by \citet{disanti01} and \citet{bro03}. Hence,
they are appropriate for the study of the origin of CO gas in
Hale-Bopp's coma.

There is observational evidence that the infrared CO
ro-vibrational lines observed by \citet{disanti01} and
\citet{bro03} in comet Hale-Bopp are optically thick. Optical
depth effects make brightness radial profiles flatter than when
optically thin conditions apply, and can lead to
misinterpretations if they are not properly taken into account. We
present a re-analysis of these lines using radiative transfer
calculations.

Both the radio and IR CO data show evidence for the presence of a
strong spiralling CO jet when comet Hale-Bopp was close to
perihelion \citep{bock+09,bro03}. This feature is considered in
the present study.

The observations of the CO radio lines are summarized in
Sect.~\ref{sec:obs}; the methods and models used for their
analysis are described in Sect.~\ref{sec:meth}, and the results
are given in Sect.~\ref{sec:res}. The analysis of the infrared
lines is presented in Sect.~\ref{sec:IR}. A summary follows in
Sect.~\ref{sec:summary}.

\section{Analysis of the CO millimetric data}

\subsection{Observations : summary} \label{sec:obs}
 The detailed description of the observations of
the CO $J$(1--0) (115 GHz) and $J$(2--1) (230 GHz) lines
undertaken in comet Hale-Bopp with the IRAM PdBI can be found in
\citet{hen03,bock+09}. They were performed on 11 March 1997 from
4.4 to 13.8 h UT. Comet Hale-Bopp was at a heliocentric distance
of 0.989 AU and a geocentric distance of 1.368 AU. Both lines were
observed simultaneously using different receivers and a spectral
resolution of 0.13 km s$^{-1}$. At that time, the PdBI comprised
five 15-m dishes. The interferometer was used in a compact
configuration C1, with baselines (the spacing between two
antennas), ranging from $\sim$ 20 to $\sim$ 150 m. Most of the
time, the comet was observed in cross-correlation
(interferometric) mode. However, every one hour short-time (1 min
on source) single-dish observations (also referred as to ON-OFF
measurements) were performed. The observing mode was
position-switching to cancel the sky background, with the
reference position taken at 5 \arcmin{} from the comet position.

When all interferometric data are considered, the full width at
half maximum (FWHM) of the synthesized interferometric beam is
2\arcsec \ $\times$ 1.38\arcsec{} at 230 GHz, and 3.58\arcsec \
$\times$ 2.57\arcsec{} \ at 115 GHz.  The FWHM of the primary beam
of the antennas is 20.9\arcsec \ at 230~GHz and 41.8\arcsec \ at
115~GHz. Given the geocentric distance of the comet, these angular
resolutions correspond to projected distances of 1650 to 41600 km
(diameter).
 Interferometric
maps of the CO $J$(1--0) and $J$(2--1) emissions are presented in
\citet{bock+09}. The signal-to-noise ratio in the center of the
maps ranges from 25 ($J$(1--0)) to 140  ($J$(2--1)). Line areas
(i.e., flux densities integrated over velocity in units of Jy km
s$^{-1}$, the lines being Doppler-resolved) measured on the
overall single-dish (ON-OFF) spectra averaging all data ($F_{\rm
SD}$), and measured at the peak of interferometric maps ($F_{\rm
Int}$), are given in Table~\ref{tab-int}. For both ON-OFF and
interferometric data, the uncertainties in flux calibration are at
most 10\% and 15\% for the $J$(1--0) and $J$(2--1) lines,
respectively \citep{bock+09}.

\subsection{Methods of analysis}
\label{sec:meth}
 The possible extended origin of CO can be
investigated by comparing single-dish $F_{\rm SD}$ and
interferometric  $F_{\rm Int}$ fluxes (Table~\ref{tab-int}).
Extended source production, if resolved out by the interferometric
beam, would result in a ratio $R$ = $F_{\rm SD}$/$F_{\rm Int}$
higher than the value expected for a pure nuclear production. On
the other hand, opacity effects need also to be carefully
investigated: being more important for smaller beams, they
increase the $F_{\rm SD}$/$F_{\rm Int}$ intensity ratio with
respect to optically thin conditions. The measured intensity
ratios $R_{\rm 1-0}$ and $R_{\rm 2-1}$ (the subscripts refer to
the lines) are given in Table~\ref{tab-int}, with error bars
including uncertainties in flux calibration.

The CO radial distribution can also be studied from the brightness
distribution observed in the interferometric maps. We preferred to
study the direct output of the interferometer, i.e., the complex
visibilities measured by the antenna pairs which sample the
Fourier transform of the brightness distribution in a number of
points in the Fourier plane (also called $uv$-plane). This
approach allows us to avoid uncertainties in map analysis related
to: 1) the ellipticity of the interferometric beam which results
from the anisotropic coverage of the Fourier plane; 2) flux losses
expected for extended sources due to the non coverage of the
$uv$-plane at baselines shorter than 20 m. Visibilities
(integrated over velocity) recorded by the ten baselines over the
whole observing period were radially averaged in the $uv$-plane
over intervals of $uv$-radius $\sigma $ of 5 m (the $uv$-radius
corresponds to the baseline length projected onto the plane of the
sky). The line area determined on the averaged single-dish
spectrum provides the visibility at $\sigma$ = 0.

The visibility from extended source emission is expected to drop
faster with increasing $\sigma$ than for more compact sources. As
shown in the Appendix, for a density distribution varying
according to $r^{-2}$, where $r$ is the distance to the nucleus,
the amplitude of the visibility $\bar{\mathcal{V}}$ is expected to
vary according to $\sigma^{-1}$, provided molecular excitation is
not varying in the coma and the line is optically thin. For an
optically thick line, the power index of the variation of
$\bar{\mathcal{V}}$ with $\sigma$ will be $<$--1. For an extended
density distribution which follows a Haser daughter distribution,
we computed that $\bar{\mathcal{V}}$ $\propto$ $\sigma^{-2}$, if
the parent scale length $L_p$ is larger than the size of the
interferometric beam and the line is optically thin. If molecular
excitation is varying in the coma (e.g., as a result of variations
of the gas kinetic temperature), $\bar{\mathcal{V}}(\sigma)$ will
deviate from these laws. The measured variation is $\propto$
$\sigma^{-1.27\pm0.02}$ and $\propto$ $\sigma^{-0.85\pm0.06}$, for
the $J$(2--1) and $J$(1--0) lines, respectively \citep[slightly
different values were derived from linear fitting by][]{bock+09}.

These two approaches were used by \citet{boi+07} to study the
radial distributions of H$_2$S, SO, and CS observed with the IRAM
PdBI.

\subsection{Modelling}
\label{sec:modelling}

Synthetic ON-OFF spectra and visibility spectra were computed
using an excitation model and a radiative transfer code, following
\citet{boi+07}. We summarize here the main aspects of the
computations.

The calculation of the population of the rotational levels of the
CO molecules considers collisions with H$_2$O and electrons, and
IR radiative pumping of the $v$(1--0) CO vibrational band
\citep{cro83,cro87,biver99b}. The model provides populations as a
function of radial distance $r$, given a H$_2$O density law with
$r$. For simplicity, we assumed an isotropic H$_2$O coma.  The
collisional CO--H$_2$O cross-section is taken equal to $\sigma_c =
2 \times 10^{-14}$ cm$^2$ \citep{biver99b}, and the H$_2$O
production rate is $Q_{\rm H_2O} = 10^{31}$~s$^{-1}$
\citep{colom99}. As shown in \citet{bock+09}, PdBI observations,
including ON-OFF measurements, were mostly sensitive to CO
molecules with rotational populations at local thermal equilibrium
(LTE). The large size of the CO LTE region can be explained by the
high collisional rates in this productive comet and the large
radiative lifetimes of the CO $J$ = 1, 2 rotational levels.

Two nominal gas kinetic temperature laws were investigated: 1) a
constant gas kinetic temperature throughout the coma equal to 120
K, which agrees with temperature determinations pertaining to the
$r \sim$ 10\,000 km coma region \citep{biver99a,disanti01}; this
temperature provides a best fit to the relative intensities of the
CO $J$(2--1) and $J$(1--0) lines measured in single-dish mode at
PdBI \citep{bock+09}, assuming pure CO nuclear production and an
isothermal coma; 2) a temperature law $T_{\rm var}$ increasing
with increasing cometocentric distance, as expected from coma
hydrodynamical models which include photolytic heating
\citep{combi99}. The long-slit infrared spectroscopic observations
of CO and other molecules undertaken in March-April 1997 in comet
Hale-Bopp suggest typical gas kinetic temperatures of $\sim$
80--90 K at $r \sim$ 1000 km \citep{magee99,disanti01}. We adopted
the temperature law used by \citet{boi+07} to interpret the PdBI
observations of sulfur-bearing species. It yields $T \leq$ 70 K at
$r \leq$ 1000 km and $T$= 120 K at $r \geq$ 10\,000 km (see
footnote in Table~\ref{tab-int-mod}). For a few calculations, we
investigate an alternate temperature law $T_{\rm var(alt)}$ with
$T_{\rm var(alt)}$ = $T_{\rm var}$ for $r \geq$ 1500 km and $T =$
79 K for $r \leq$ 1500 km.

\begin{figure}
\includegraphics[width= 13cm]{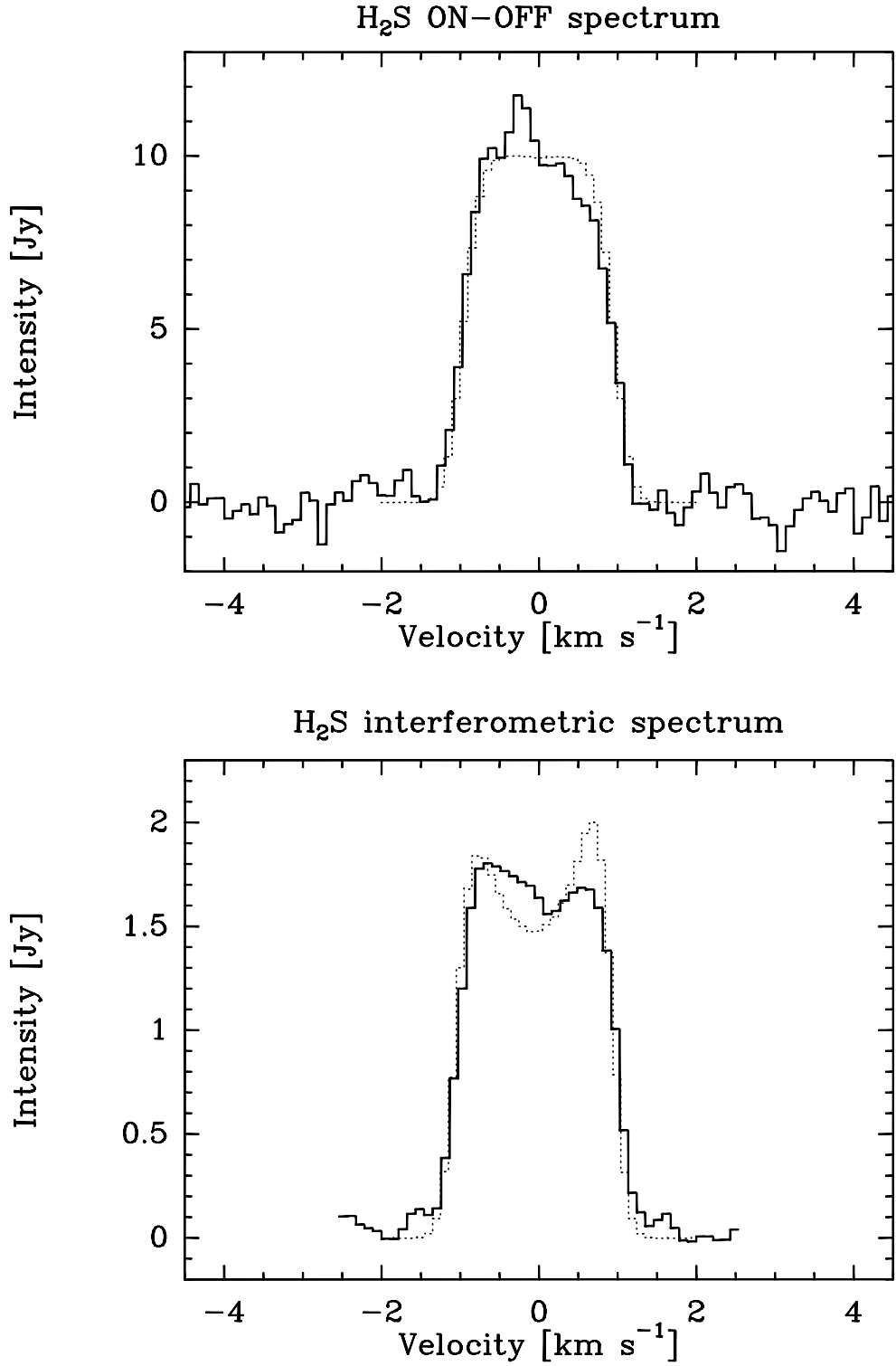}
\caption{ON-OFF and interferometric spectra of the H$_2$S
2$_{20}$--2$_{11}$ line at 216.7 GHz obtained in comet Hale-Bopp
on 13 March 1997 with the PdBI \citep{boi+07}. Synthetic spectra
computed using the $v_{var}$ law for the expansion velocity are
shown with dotted lines. The angular resolution is 22.2\arcsec~and
1.7\arcsec, for the ON-OFF and interferometric beams,
respectively.} \label{h2s-sp}
\end{figure}

In most calculations, the CO density distribution was computed
using simple steady state isotropic models. We considered outflow
either at a constant velocity (i.e., Haser models) or at a
velocity which increases with increasing cometocentric distance.
Indeed, gas acceleration is expected in the $r$ = 1000--10\,000 km
region probed by the PdBI in the Hale-Bopp coma, as a result of
photolytic processes \citep{combi99}. The shape of radio lines
provides constraints on the gas expansion velocity $v_{\rm exp}$
\citep[e.g., ][]{hu91}. From the width of radio lines of parent
molecules (including CO) observed by single-dishes,
\citet{biver99a} derived a value equal to 1.05 km s$^{-1}$ for the
period pertaining to the PdBI CO observations. This value is
representative of the gas velocity at cometocentric distances of
typically the radius of the field of view and therefore pertains
to radial distances $\sim$ 10\,000 km. The width of the CO lines
in the interferometric beam suggests that $v_{\rm exp}$ $\sim$ 0.9
km s$^{-1}$ at $r$ $\sim$ 1000 km. For comparison, the model of
\citet{combi99} yields to an acceleration from 0.91 to 1.12 km
s$^{-1}$ in the $r$  = 1000--10\,000 km region, in good agreement
with our velocity retrievals. For the models with constant
velocity ($v_{\rm const}$), we adopted $v_{\rm exp}$ = 1.05 km
s$^{-1}$. For the models which consider the variation of the
velocity ($v_{\rm var}$), we assumed $v_{\rm exp} = 0.9+0.15
\times \log(r[\rm km]/1000)$ km s$^{-1}$ for $r > 1000 $ km,
$v_{\rm exp}$ = 0.9 km s$^{-1}$ for $r < 1000$ km.
Figure~\ref{h2s-sp} shows the H$_2$S 2$_{20}$--2$_{11}$ line at
216.7 GHz observed in comet Hale-Bopp on 13 March 1997 with the
PdBI \citep{boi+07}. Synthetic line profiles computed with the
$v_{\rm var}$ velocity law are superimposed. The width and wings
of both the ON-OFF and interferometric spectra are reproduced. The
spectral asymmetries present in the central part of the observed
H$_2$S spectra are due to spatial asymmetries in the coma. CO and
H$_2$S are expected to expand at the same expansion velocity. We
checked that the $v_{\rm var}$ law consistently explains the width
and wings of the CO $J$(1--0) and $J$(2--1) line profiles.

In theory, cometary line shapes contain information on the radial
distribution of the molecules. Daughter molecules are expected to
exhibit double-horn line shapes when the size of the beam does not
exceed much their production scale length. The missing flux in the
center of the line depends on the ratio of these two quantities.
However in comet Hale-Bopp, CO line profiles were found to be
affected by the presence of a CO rotating jet which introduced
excess emission alternatively in the blue and red part of the
profiles during the rotation \citep{bock+09,boi+09}. Such
complexity prevents detailed fitting of the CO line shapes.

Extended distributions were investigated using the Haser formula
for daughter species with the parent and daughter scale lengths
$L_p$ and $L_d$ and the gas velocity $v_{\rm exp}$ as parameters.
For simplicity, both $L_p$ and $L_d$ (= $L_{\rm CO}$) were assumed
to be constant throughout the coma, i.e., we did not consider the
effect of the variation of the expansion velocity on the scale
lengths. Therefore, in all calculations we assumed $L_{\rm CO}$ =
$v_{\rm exp}/ \beta_{\rm CO}$, with $v_{\rm exp}$ = 1.05 km
s$^{-1}$ and $\beta_{\rm CO} = 7.5 \times 10^{-7} s^{-1}$
\citep{huebner92}. Such approximation has a negligible effect on
the results. Indeed, the CO scale length is much larger than the
field of view of the single-dish measurements, so the CO column
density within the field of view is not much affected by small
inaccuracies in the effective CO scale length. The relative
contribution of the extended and nuclear productions of CO is
given by the ratio of the production rates $Q_{\rm CO }^N$:$Q_{\rm
CO}^E$, where the subscripts $N$ and $E$ refer to the nuclear and
extended productions, respectively. The total production rate is
$Q_{\rm CO }$=$Q_{\rm CO }^N$+$Q_{\rm CO}^E$.

We also consider in Sect.~\ref{mod-jet} much complex distributions
where a large fraction of the CO production is within a spiralling
jet. As mentioned above, a strong CO jet was indeed identified in
the coma of comet Hale-Bopp from these interferometric
observations. The CO distribution is modelled using
tridimensional, time-dependent codes \citep{bock+09,boi+09}.

The brightness distribution of the two CO lines was computed using
the radiative transfer code of \citet{boi+07}, which includes both
self-absorption and stimulated emission \citep[see][for detailed
explanations]{biv97}. ON-OFF synthetic spectra were obtained by
convolving the brightness distributions with the primary beam
pattern of the PdBI antennas, described by a 2D Gaussian.
Visibilities were calculated from the Fourier transform of the
brightness distribution \citep{boi+07}. Maps of the CO line
emissions were synthesized, considering the coverage of the
Fourier plane obtained during the observations. The peak flux
densities in the modelled and observed maps, both obtained by
inverse Fourier Transform, can then be directly compared.

\section{The CO radial distribution from millimetric data}
\label{sec:res}
\subsection{ON-OFF and interferometric flux densities}

Table~\ref{tab-int-mod} gives flux densities and intensity ratios
$R_{Mod}$ = $F_{\rm SD}$/$F_{\rm Int}$ computed with $Q_{\rm CO}$
= 2.1 $\times$10$^{30}$ s$^{-1}$ and several Haser distributions
describing the local CO density. Except for the first entry,
corresponding to optically thin assumptions, all results were
obtained with radiation transfer calculations.

The value $Q_{\rm CO}$ =  2.1 $\times$10$^{30}$ s$^{-1}$ was
chosen so that the $J$(1--0) and $J$(2--1) ON-OFF line intensities
computed with $T$ = 120 K, a constant velocity and a pure CO
nuclear production match precisely the measurements. This value is
consistent with the total CO production rate (sum of native and
extended distributions) inferred from IR spectroscopy
\citep{disanti01}. The other models considered in
Table~\ref{tab-int-mod} yield ON-OFF line intensities consistent
with the measured values if one takes flux uncertainties into
account.

To quantify the goodness of the models in explaining the observed
intensity ratios $R_{\rm Obs}$ ($R_{1-0}$ or $R_{2-1}$), we
computed for both lines the quantity:
\begin{equation}
\chi^2 = (R_{\rm Obs}-R_{\rm Mod})^2/\Delta R_{\rm Obs}^2
\end{equation}
where $\Delta R_{\rm Obs}$ is the uncertainty on $R_{\rm Obs}$
given in Table~\ref{tab-int} and $R_{\rm Mod}$ is the intensity
ratio given by the models. Results are plotted in
Fig.~\ref{plot-ratio}. Models resulting in $\chi^2$ $>$ 1 in
either line are considered as non satisfactory.

\subsubsection{Models with pure nuclear production}

\begin{figure}
\resizebox{\hsize}{!} {
\includegraphics[angle=270,width=13cm, bb = 100 50 550 530]{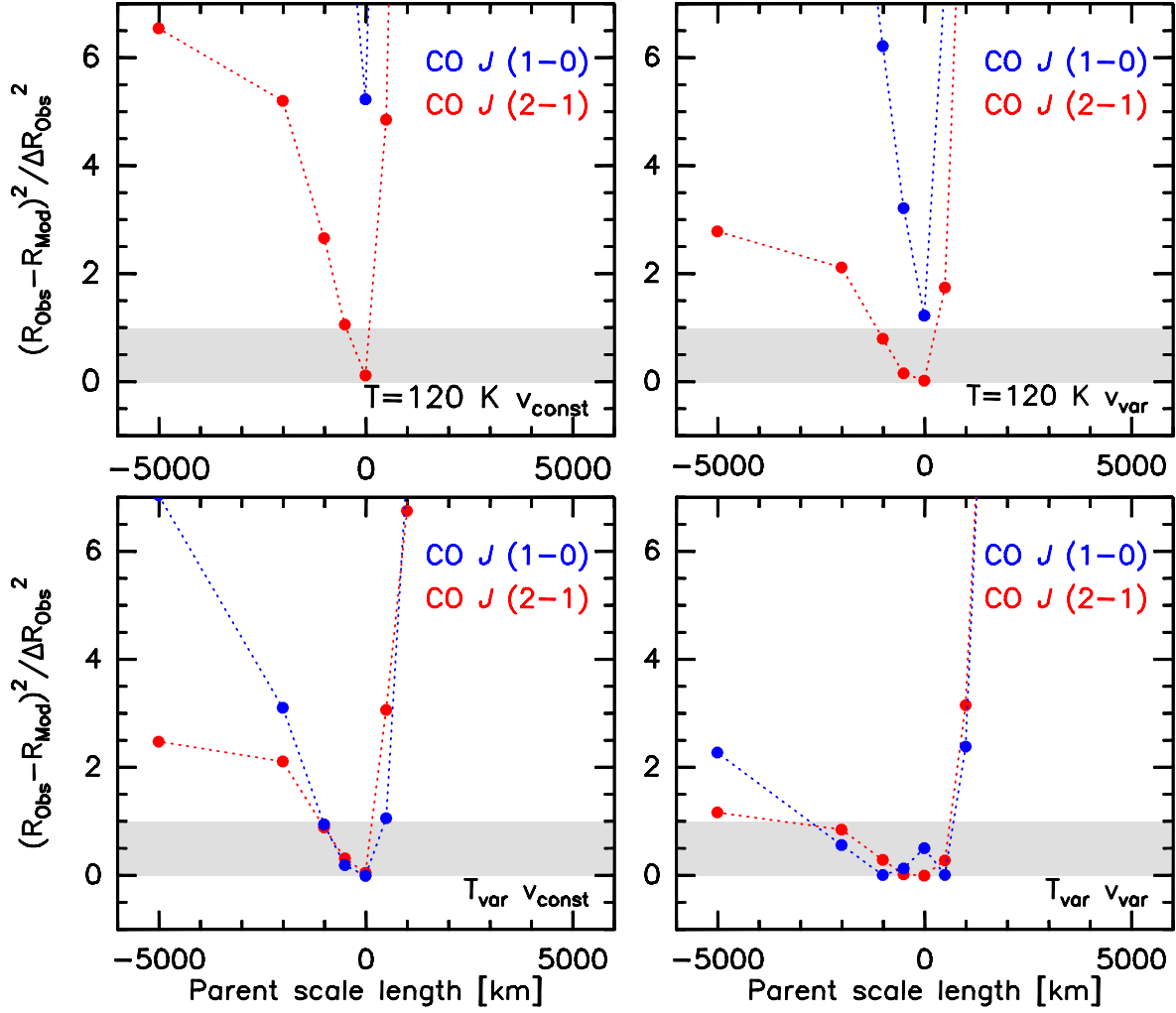}
}
 \caption{$\chi^2$ = $(R_{\rm
Obs}-R_{\rm Mod})^2/\Delta R_{\rm Obs}^2$ as a function of the CO
parent scale length, where $R_{\rm Obs}$ and $R_{\rm Mod}$ are the
observed and modelled intensity ratios $R$= $F_{\rm SD}$/$F_{\rm
Int}$, respectively, and $\Delta R_{\rm Obs}$ is the error bar on
$R_{\rm Obs}$. Blue (respectively red) symbols are for the CO
$J$(1--0) (respectively $J$(2--1)) lines. Results plotted in the
right part of the plots (positive scale lengths) are for pure
extended distributions (i.e., $Q_{\rm CO }^N$:$Q_{\rm CO }^E$ =
(0:100)). Results for composite distributions ($Q_{\rm CO
}^N$:$Q_{\rm CO }^E$ = (50:50)) are shown in the left part of the
plot (negative scale lengths). The grey region corresponds to 0
$<$ $\chi^2$ $<$ 1. The four quadrants correspond to different gas
temperature and velocity laws indicated in the bottom right
corner.} \label{plot-ratio}
\end{figure}

The $J$(2--1) line observations can be fully interpreted by a
nuclear CO production, provided the opacity effects are taken into
account.  Opacity affects mainly the flux density of the $J$(2--1)
line in the interferometric beam: e.g., a reduction by 37\% is
observed for the model with constant temperature and velocity,
while ON-OFF $J$(2--1) flux and $J$(1--0) fluxes are almost
unchanged (compare the first two entries in
Table~\ref{tab-int-mod}). When line opacity is considered, the
$R_{2-1}$ values obtained with either of the temperature and
velocity laws ($\sim$ 18--19) are consistent with the measured
value of 17.8 $\pm$ 3.8. $\chi^2$ values close to zero are
obtained (Fig.~\ref{plot-ratio}).

On the other hand, models considering a nuclear production and a
constant gas kinetic temperature fail to explain the ratio
$R_{1-0}$ = 11.6$\pm$1.8 measured for the $J$(1--0) line, yielding
values 20--35\% higher than observed. The discrepancy obtained for
$R_{1-0}$ cannot be solved by invoking an extended CO production
because this process affects the intensity ratio in the opposite
way (Fig.~\ref{plot-ratio}). However, a good agreement is obtained
when the variable temperature law is used: the population of the
$J$ = 1 CO rotational level increases with decreasing $T$, hence
producing an enhancement of the $J$(1--0) emission from the inner
colder parts. The $R_{2-1}$ ratio is less sensitive to the assumed
gas temperature law because the increase of $J$(2--1) local
emission in the inner regions is counterbalanced by more
significant self-absorption effects for the emergent emission.

The $R_{1-0}$ and  $R_{2-1}$ ratios are both sensitive to the
assumed velocity law which affects the CO density radial profile.
$R_{1-0}$ decreases when the expansion velocity in the inner coma
decreases, the outflow velocity at $r$ $>$ 10\,000 km being kept
constant. $R_{2-1}$ displays a non monotonous trend with
decreasing inner coma velocity $v_{\rm exp}$(inner), slowly
decreasing for 0.9 km s$^{-1}$ $<$ $v_{\rm exp}$(inner) $<$ 1.05
km s$^{-1}$ (Table~\ref{tab-int-mod}), and then increasing for
$v_{\rm exp}$(inner) $<$ 0.9 km s$^{-1}$. We found that it is not
possible to explain both $R_{1-0}$ and $R_{2-1}$ measured ratios
with a variable velocity law and a constant gas kinetic
temperature.

In summary, the measured intensity ratios $R_{1-0}$ and $R_{2-1}$
can be explained with a CO nuclear origin when the variation of
the gas kinetic temperature with distance to nucleus is
considered.

\subsubsection{Models with extended CO production}

Let us first compare the measurements to model results obtained
with pure CO daughter Haser distributions. Indeed, \citet{bro03}
interpreted long-slit infrared observations of CO in comet
Hale-Bopp, with $\sim$ 90\% of the CO being released from a source
with $L_p$ = 5000 $\pm$ 3000 km. We investigated CO parent scale
lengths up to $L_p$ = 5000 km. Such CO distributions with $L_p$
$>$ 500 km do not fit the data (Fig.~\ref{plot-ratio}). In other
words, both the CO $J$(1--0) and $J$(2--1) data exclude that CO is
produced solely by an extended source, unless this source releases
most of the CO gas in the near-nucleus coma unresolved by the
interferometric beam. Results obtained with $Q_{\rm CO
}^N$:$Q_{\rm CO}^E$ in ratio 10:90 and $L_p$ = 5000 km do not
reproduce the measurements (Table~\ref{tab-int-mod}), and thereby
do not support the conclusions of \citet{bro03}.

\citet{disanti01} interpreted the radial profiles of CO infrared
emission by the contribution of a distributed source to 50 \% of
the total CO production. Therefore, we also investigated composite
nuclear:extended productions with CO and CO parent production
rates in ratio 50:50 (Table~\ref{tab-int-mod}, and left part of
the plots shown in Fig.~\ref{plot-ratio}). Satisfactory fits to
both lines can only be obtained for models with the variable
temperature law: the $R_{1-0}$ and $R_{2-1}$ ratios can be
reproduced within 1-$\sigma$ providing $L_p$ $<$ 1000 km (model
with constant velocity) or $L_p$ $<$ 2000 km (model $v_{\rm
var}$). One can consider that the condition
$\chi^2_{1-0}+\chi^2_{2-1}$ $<$ 1 should be satisfied as the
$J$(1--0) and $J$(2--1) measurements are independent. Then, models
with $L_p$ $<$ 1500 km ($v_{\rm var}$ law) and $L_p$ $<$ 800 km
($v_{\rm const}$ law) provide a satisfactory fit to the PdBI CO
data based on the ratios $R$ of single-dish to interferometric
fluxes. Since the adopted $v_{\rm var}$ law is consistent both
with line shape measurements and gas dynamics calculations
(Sect.\ref{sec:modelling}), we conclude that composite 50:50
nuclear/extended productions with $L_p$ $>$ 1500 km are excluded
by the CO millimeter data.

It is worth noting that, using the models with the $v_{\rm var}$
and $T_{\rm var}$ laws, and extended productions fitting the data,
we infer total CO production rates within 5\% of the nominal value
of 2.1 $\times$10$^{30}$ s$^{-1}$.

Still, we assumed isotropic CO density distributions. In
Sect.~\ref{mod-jet}, we discuss how the $F_{\rm SD}$/$F_{\rm Int}$
ratio is affected by the presence of jets.

\subsection{Radial evolution of the visibilities}

\begin{figure}
\begin{center}
\includegraphics[angle=270, width=11cm]{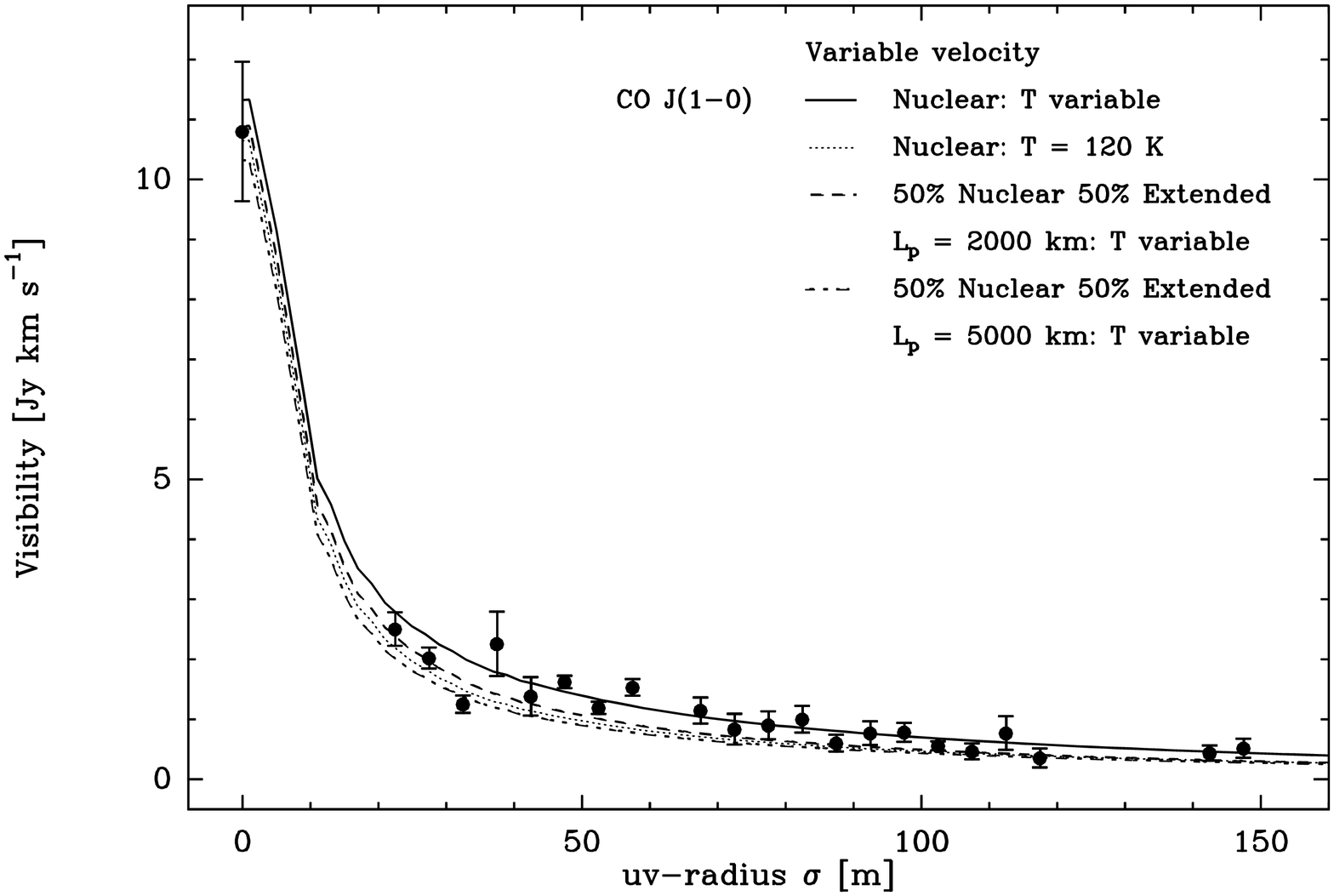}
\includegraphics[angle=270, width=11cm]{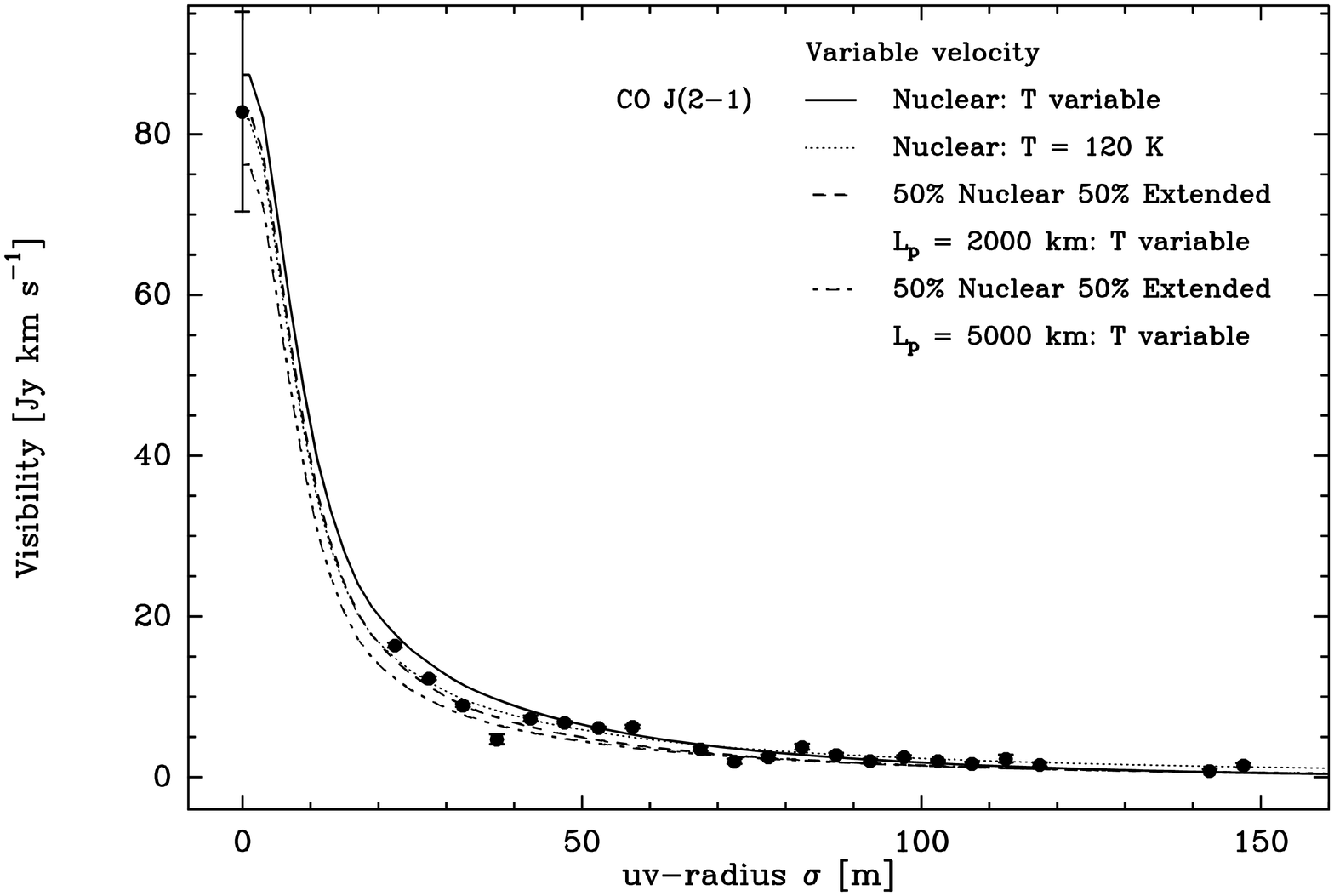}
\end{center}
\caption{Amplitude of the visibility as a function of $uv$-radius
for the CO $J$(1--0) line (top) and the CO $J$(2--1) line
(bottom). Measurements are shown by dots with error bars which
correspond to statistical noise uncertainties, except for the
ON-OFF measurements where calibration uncertainties (10\% and 15\%
for $J$(1--0) and $J$(2--1), respectively) are added
quadratically. The different curves correspond to different
models: nuclear production with variable gas temperature in the
coma (plain line), nuclear production with constant temperature
(dotted line), combined nuclear/extended productions in ratio
50:50 with $L_p$ = 2000 km and variable temperature (dashed line),
combined nuclear/extended productions in ratio 50:50 with $L_p$ =
5000 km and variable temperature (dotted-dashed line). All models
consider isotropic outgassing at a velocity described by $v_{\rm
var}$ and a total CO production rate $Q_{\rm CO }$ = 2.1
$\times$10$^{30}$ s$^{-1}$.} \label{visi-rad}
\end{figure}

\begin{figure}
\begin{center}
\includegraphics[angle=270,width=11cm]{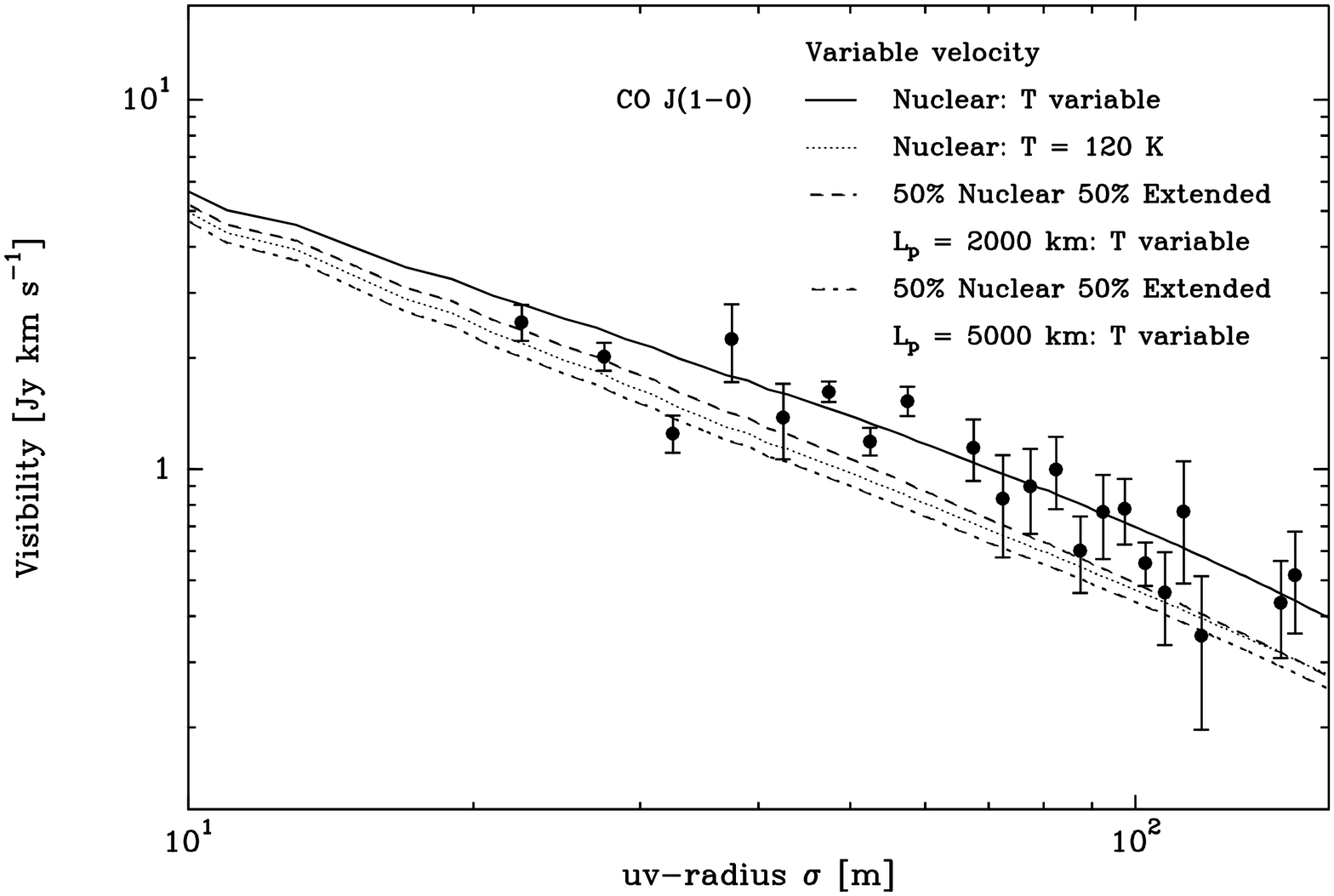}
\includegraphics[angle=270,width=11cm]{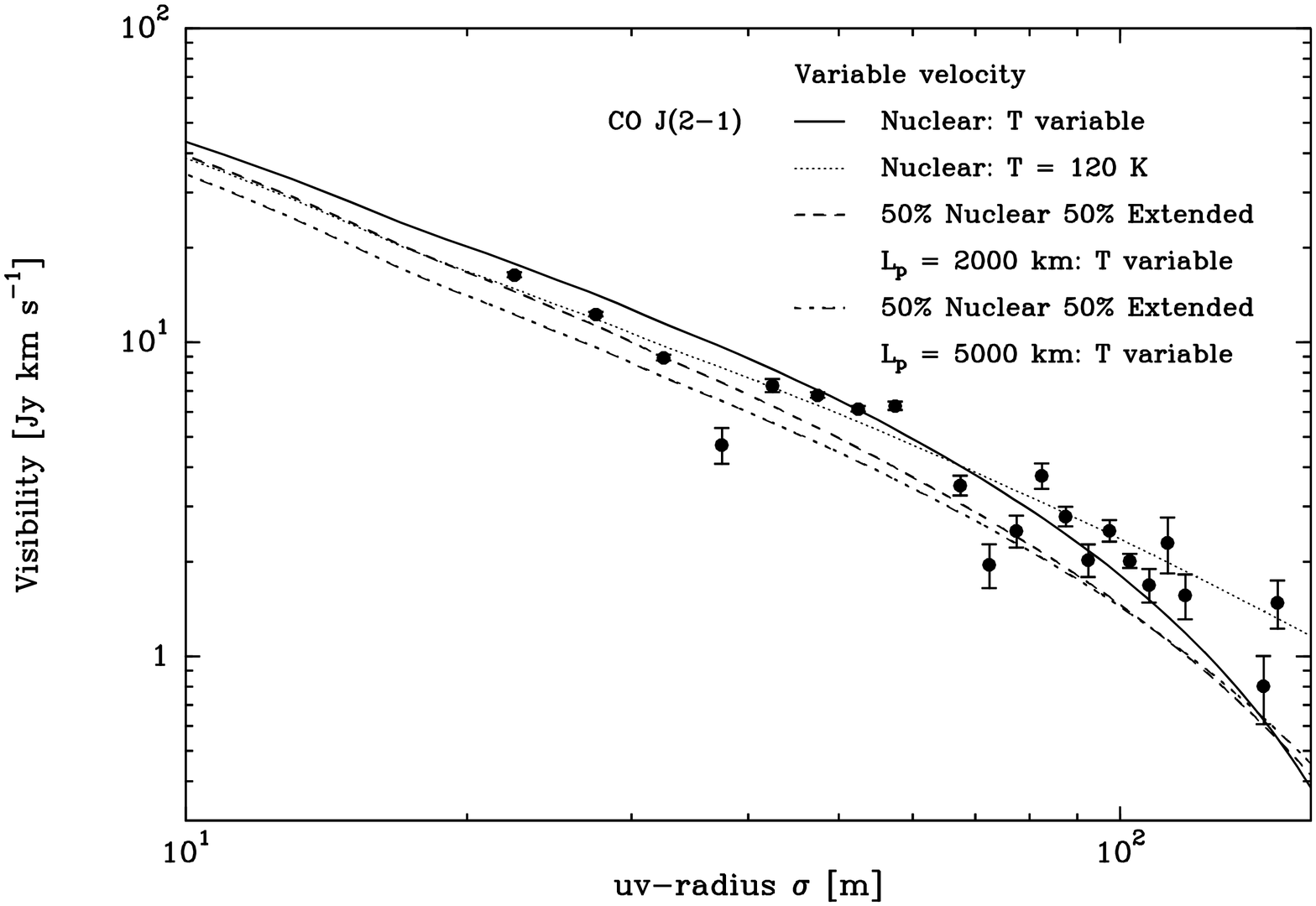}\end{center}
 \caption{Log-Log
plot of the amplitude of the visibility as a function of
$uv$-radius for the CO $J$(1--0) line (top) and the CO $J$(2--1)
line (bottom). Model parameters are the same as for Fig.
\ref{visi-rad}.} \label{visi-rad-log}
\end{figure}

In Figs~\ref{visi-rad}--\ref{visi-rad-log} are plotted the
visibilities $\bar{\mathcal{V}}$ as a function of $uv$-radius
$\sigma$ expected for different models, together with the
measurements. The power index of the variation of
$\bar{\mathcal{V}}(\sigma)$  with  $\sigma$ measured on the
modelled curves is given in Table \ref{tab-int-mod}. The
least-squares fit to the computed visibilities was performed on an
extracted set of visibilities representative of the $\sigma$
sampling of the observations. Weighting factors corresponding to
the uncertainties in the measured visibilities were considered
when fitting the extracted set of computed visibilities. This
method is appropriate for best comparison with the data, since the
model results show that the visibility curves do not follow a
simple power law (Fig.~\ref{visi-rad-log}).

From visual comparison between the modelled and observed
visibility curves (Figs~\ref{visi-rad}--\ref{visi-rad-log}), we
see that :

\begin{enumerate}

\item Both lines exclude pure CO production from an extended
source, unless the characteristic parent scale length of the
distributed source is well below 500 km;

\item Composite nuclear/extended productions with $L_{\rm p}$ $=$
5000 km or 2000 km do not explain the observations.

\item The best fits are obtained for a nuclear production.

\end{enumerate}

 From a least square fitting of the visibility curves, including
the ON-OFF point, we found that composite 50:50 nuclear/extended
productions with $L_{\rm p}$ $>$ 1500 km are not statistically
favored when the $v_{\rm var}$ law is considered. In other words,
we obtained the same results as derived from the intensity ratios.

\begin{figure}
\begin{center}
\includegraphics[angle=270, width=\textwidth]{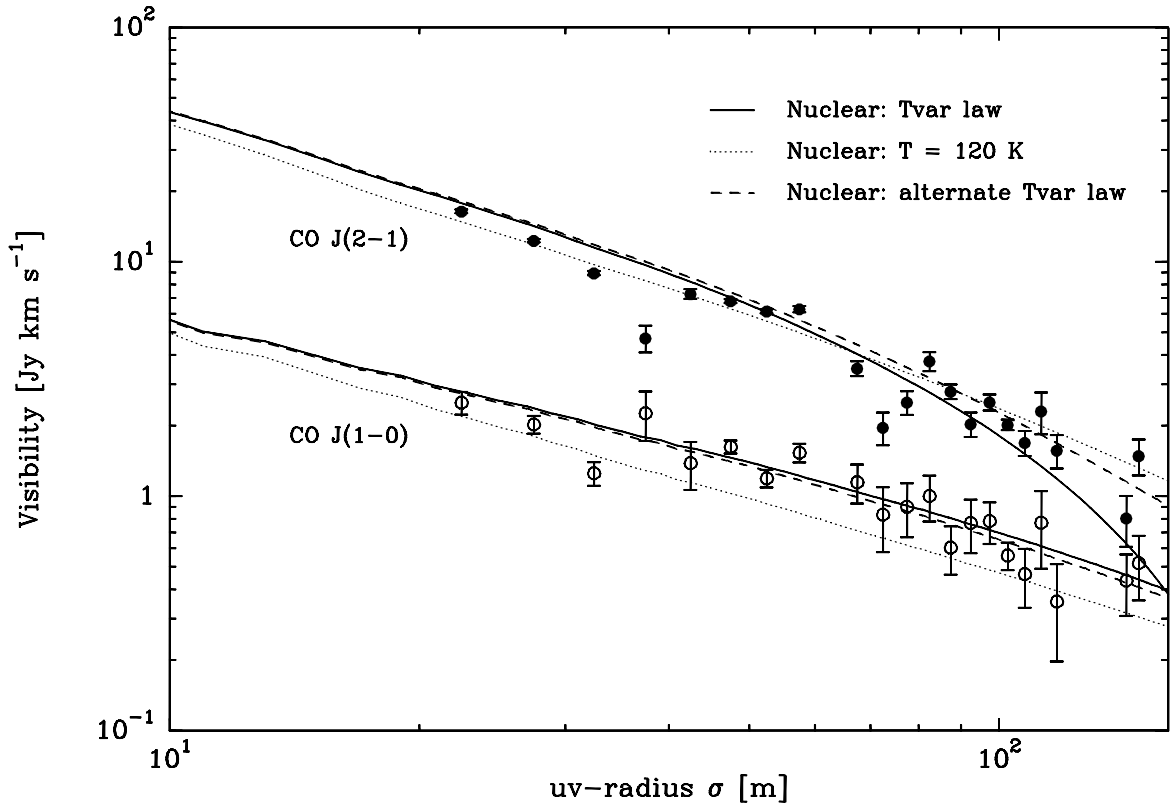}
 \end{center}
 \caption{Amplitude
of the visibility as a function of $uv$-radius. CO $J$(1--0)
(respectively $J$(2--1)) measurements are shown with open
(respectively filled) symbols with error bars. The different
curves correspond to calculations for a CO nuclear production and
different temperature laws. Other model parameters are the same as
for Fig. \ref{visi-rad}.} \label{visi-rad-t}
\end{figure}

\begin{figure}
\begin{center}
\includegraphics[angle=270,width=13cm, bb = 260 220 480 490]{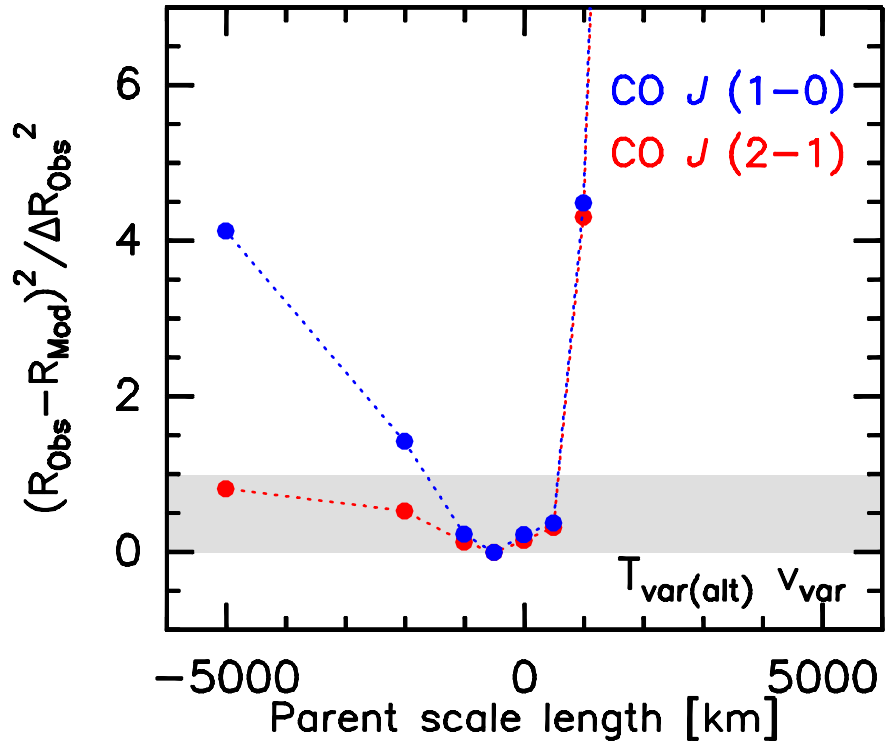}
\end{center}
 \caption{Same as Fig.~\ref{plot-ratio}, using the $T_{var(alt)}$ and
 $v_{var}$ laws}. \label{plot-ratio-alt}
\end{figure}

Let us now compare the power indexes of modelled visibility curves
to the measured values. For the $J$(1--0) line, there is a good
agreement (within 1.3-sigma) between the measured value and that
given by the model with a nuclear CO production, and $T_{\rm var}$
and $v_{\rm var}$ laws. On the other hand, the same model predicts
a power index of $-1.40$ for $J$(2--1) whereas the measured value
is $-1.27\pm0.02$. The steep decrease of the $J$(2--1) visibility
curve obtained with the $T_{\rm var}$ law at large $uv$-radii ($>$
70 m, see Fig.~\ref{visi-rad-log}) is caused by large
self-absorption effects in the inner cold coma observed by the
interferometric beam ($r < 850$ km) where $T$ $<$ 65 K. At 230
GHz, $uv$-radii between 70--150 m probe cometocentric distances
($\propto \lambda/\sigma$) to which 115 GHz measurements are not
sensitive. Unlike $J$(2--1) visibilities, $J$(1--0) visibilities
are not influenced by the coma temperature at $r$ $<$ 1500 km.
Figure~\ref{visi-rad-t} shows that the $J$(2--1) data at large
$uv$-radii can be better fitted using the alternate temperature
law $T_{var(alt)}$ introduced in Sect.~\ref{sec:modelling} for
which $T$ = 79 K at $r$ $<$ 1500 km and $T$ = $T_{var}$ at $r$ $>$
1500 km. The power index of the visibility curves computed with
this alternate temperature law are $-1.35$ and $-0.98$ for
$J$(2--1) and $J$(2--1) (Table~\ref{tab-int-mod}), respectively,
in reasonable agreement with the measurements. We plot in
Fig.~\ref{plot-ratio-alt} curves of $\chi^2$ values obtained with
the $T_{var(alt)}$ law, which are analogous to those shown in
Fig.~\ref{plot-ratio}.

The model with composite 50:50 nuclear/extended production and
$L_p$ $=$ 5000 km closely reproduces the $J$(2--1) measured power
index, but not the flux ratio $R_{2-1}$ (Table~\ref{tab-int-mod}).
It is unable to explain either the power index of the $J$(1--0)
visibility curve and $R_{1-0}$. This shows that best constraints
on the radial distribution of CO are obtained from the comparison
of ON-OFF and interferometric flux measurements.


\subsection{Models with anisotropic CO production}
\label{mod-jet}

A question that may arise is the extent to which our approach is
appropriate in presence of jets in the coma. A spiralling CO jet
comprising $\sim$ 30 \% of the total CO production was identified
from these PdBI data \citep{bock+09}. This jet, found to be issued
from a low latitude nucleus region and to be $\sim$ 40\degre wide,
caused temporal modulations in the visibilities recorded by each
pair of antennas, which are still visible in the radial and time
averages shown in Fig.~\ref{visi-rad} and \ref{visi-rad-log}.

The question has been thoroughly studied for spiralling jets of
various strengths and latitudes in the assumption of negligible
optical depth and a constant coma temperature \citep{hen03}.
\citet{hen03} used a simple description of the Hale-Bopp CO coma
represented by the combination of an isotropic coma and a
spiralling conical jet. These 3-D time dependent simulations show
that, for nearly equatorial jets, the power index of the
visibility curve remains close to $-1$, except for collimated jets
where the slope can reach $-0.8$ to $-0.7$ depending on the jet
strength. An opposite trend is observed for high-latitude jets
\citep[see results obtained for optically thin conditions in Figs
20, 21 of][]{bock+09}. These trends are explained by the sampling
time of the measurements (2/3 of the nucleus rotation, i.e.,
almost one full rotation), the orientation of the different
baselines with respect to the jet direction, and the shape of the
modulations. The shortest baselines probed more particularly gas
emissions along the rotation axis, while the longest baselines
were sensitive to equatorial jets \citep[see Figs 18
of][]{bock+09}. Hence, in presence of collimated equatorial jets,
the slope of the visibility curve decreases. Simulations were not
performed  for the $J$(1--0) line in presence of a CO jet
\citep{hen03,bock+09}. However, we do not expect conclusions
different from those obtained for $J$(2--1) since $J$(1--0) and
$J$(2--1) were observed simultaneously, i.e., with the same
baseline orientations. Less contrasted visibility modulations and
slope variations are even expected for $J$(1--0) in presence of
jets, since the spatial resolution is degraded at 115 GHz.

In the simulations of the CO PdBI data performed by
\citet{bock+09}, optical depth effects are considered and the
kinetic temperature is assumed to be equal to 120 K. When the jet
parameters that provide the best fit to the data are considered
(jet model (3)), the power index of the $J$(2--1) visibility curve
agrees with that measured on the observed visibilities
\citep{bock+09}.

We also investigated to which extent the flux ratio $R_{\rm 21}$
is affected by the presence of the CO jet. A value of 18.4,
consistent with observations, is derived from the synthesized
interferometric maps obtained by \citet{bock+09} with their jet
model (3). Other model simulations are worth to consider. In
contrast to the simple description of the coma made by
\citet{hen03} and \citet{bock+09}, \citet{boi+09} simulated the CO
PdBI observations of comet Hale-Bopp using a tridimensional
time-dependent gas-dynamical model of the coma. An increased CO
production from a localized region on the nucleus surface is
considered to produce the observed CO spiralling structure.
$R_{\rm 21}$ values in the range 18--18.6 are inferred from their
solutions 3--5 which provide the most satisfactory fits to the
data. These values are comparable to the values found under the
assumption of isotropic CO production (Table~\ref{tab-int-mod})
and to the observed value.


\section{The CO radial distribution from infrared data}
\label{sec:IR}

\subsection{General considerations}

The millimetric and infrared CO data lead apparently to
inconsistent results concerning the origin of CO in Hale-Bopp's
coma. Based on radiative transfer calculations of CO $v$ = 1--0
ro-vibrational line emission at 4.7 $\mu$m, we argue that the
analyses of \citet{disanti01} and \citet{bro03} were hampered by
an improper account of the opacity of these lines.

Qualitatively, opacity effects are an attractive explanation:

\begin{itemize} \item Being more important
for the central pixels aimed at the inner coma, they make line
brightness radial profiles to be flatter than when optically thin
conditions apply, and to mimic profiles for species with extended
production;

\item They are all the more important since the CO production rate
is large and the geocentric distance is low. \citet{disanti01}
found that in comet Hale-Bopp the CO extended source was present
for $r_h$ $<$ 1.5 AU ($Q_{\rm CO}$/$\Delta$ in the range 5--16
$\times$ 10$^{29}$ s$^{-1}$AU$^{-1}$) and inactive for $r_h$ $>$ 2
AU ($Q_{\rm CO}$/$\Delta$ $<$ 1 $\times$ 10$^{29}$
s$^{-1}$AU$^{-1}$). Interestingly, comet Hyakutake was also a
CO-rich comet and the CO infrared data showed extended brightness
profiles when $Q_{\rm CO}$/$\Delta$ was 1 to 5 10$^{29}$
s$^{-1}$AU$^{-1}$ \citep{disanti03}. Since the optical thickness
increases with decreasing gas kinetic temperature, one expects
opacity effects to be relatively more prominent in the CO
brightness profiles of comet Hyakutake where the gas temperature
was measured to be smaller. We are not aware of any other comet
observed by means of high-resolution infrared spectroscopy showing
evidence of an extended CO distribution \citep{disanti09}. So far,
for all other comets studied to date in the infrared, $Q_{\rm
CO}$/$\Delta$ was much lower than 10$^{29}$ s$^{-1}$AU$^{-1}$.

\item Comparison of the intensity of lines emitted by a common
ro-vibrational level provides direct evidence for significant
opacity effects in the lines of sight passing close to the
nucleus. Under optically thin conditions, the ratios of P2 to R0
(upper rotational level $J_u$ = 1), and of P3 to R1 ($J_u$ = 2)
line intensities are equal to 2 and 3/2, respectively (i.e.
($J_u$+1)/$J_u$). Smaller values are observed, consistent with the
higher absorption coefficients of the P2 and P3 lines as compared
with those of R0 and R1, respectively \citep{disanti01}. The
radial brightness profile of R0 is less extended than those of R1,
R2 and R3, which can be explained by the lower R0 absorption
coefficient \citep[see Fig. 13 of][]{bro03}.

\end{itemize}

Opacity effects affect the fluorescence excitation of the CO $v$ =
1 vibrational band in the inner coma, the solar photons being
absorbed along their path. Opacity effects also result in
self-absorption of infrared photons emitted by nearby CO molecules
(radiation trapping), which counterbalances the reduced direct
excitation by solar photons. Last, but not least, the received
radiation can be severely decreased by self-absorption effects.
\citet{disanti01} considered the attenuation of the solar pump and
computed effective g-factors accounting for optical depth effects.
They did not studied quantitatively the fate of emitted photons,
and assumed optically thin lines. Using effective g-factors, their
conclusions remained essentially unchanged concerning the extended
source of CO in comet Hale-Bopp. \citet{disanti03} found that the
attenuation of the solar pump explains in large part the CO
extended brightness profiles observed in comet Hyakutake,
especially those obtained end of March 1996 at $\Delta$ = 0.1 AU.
\citet{bro03} considered optical effects both in the solar pump
and in the emitted radiation to explain data obtained on 5 March
1997, and found spatial profiles only slightly wider than in the
optically thin case.

We note that, despite CO being less abundant than water, its $v$ =
1--0 ro-vibrational lines are expected to be much thicker than the
H$_2$O lines from hot bands observed in the infrared. Photons
emitted from a ro-vibrational transition ($v$',
i)$\rightarrow$($v$'', j) can only be re-absorbed by the ($v$'',
j) level to excite ($v$', i). In cometary atmospheres, most of the
molecules are in their ground fundamental vibrational state.
Hence, self-absorption is much less important for H$_2$O hot
bands, which connect weakly populated high vibrational states,
than for the CO $v$ = 1--0 band. Using the infrared excitation
model of \citet{bock89}, we computed that the opacity of
 $\nu_3$--$\nu_2$ H$_2$O lines observed in comet Hale-Bopp \citep{dello00,bro03} is
typically 5-6 orders of magnitude lower than the opacity of CO $v$
= 1--0 lines at 1 AU from the Sun.  For example, we can compare
the P3 ($J$ = 2 $\rightarrow$ 3) line of CO $v$ = 1--0 (2131.6
cm$^{-1}$) to the $2_{12}$ $\rightarrow$ $3_{03}$ line of H$_2$O
$\nu_3$--$\nu_2$ (2003 cm$^{-1}$). The line absorption
coefficient, and therefore the opacity, is proportional to the
Einstein coefficient for absorption $B_{ji}$ ($\propto$ $A_{ij}$,
the Einstein coefficient for spontaneous emission) times the
population of the lower level of the transition $n_j$. Assuming
that the rotational levels of the ground vibrational state are
populated according to a Boltzmann distribution at 80 K, for the
CO line we have $A_{ij}$ = 20.3 s$^{-1}$ and $n_j$ = 0.14, while
for the H$_2$O line $A_{ij}$ = 1.14 s$^{-1}$ and $n_j$ = 7.0
10$^{-6}$.

\subsection{Modelling}

We performed detailed calculations of CO $v$ = 1--0 line
intensities solving exactly the radiative transfer by integration
through the coma. The source functions at each radial distance
were determined using the escape probability formalism
\citep{dbm87}. This method allows one to solve the statistical
equilibrium equations for the level populations decoupled from any
radiative transfer. It accounts for absorption of nearby emitted
photons, stimulated emission, and for the attenuation of the solar
flux by a correction factor to the excitation rates. \citet{zak07}
showed that the escape probability formalism is appropriate to
account for radiation trapping in cometary atmospheres. Escape
probabilities were computed assuming spherical expansion at
constant velocity \citep{dbm87}. The same approach was used by
\citet{bro03}. To compute fluorescence excitation of the $v$ = 1
vibrational band, the population distribution in the ground
vibrational state is needed. We used the excitation model for
rotational emission described in Sect.\ref{sec:meth}. For
completeness, collisional quenching of $v$ = 1--0 fluorescence was
considered assuming a collisional cross-section for de-excitation
equal to 2.5 $\times$ 10$^{-16}$ cm$^2$, H$_2$O being the
collision partner.  However, this process was found unimportant
for comet Hale-Bopp close to perihelion, except within a hundred
of kilometers from the nucleus surface. Because of Doppler shifts,
the optical depth experienced by the CO emitted photons depends on
the velocity distribution and radial profile. As for the radiative
transfer calculations of CO rotational line emission, constant or
variable velocity radial profiles were assumed. The local velocity
dispersion was described by thermal broadening.

Two radiative transfer codes were developed. In the first one,
which results are presented in Sect.\ref{sec:Qcurve-iso},
isotropic Haser distributions were considered. The second one is
an adaptation of the code of \citet{bock+09} used to interpret the
CO interferometric radio data (see Sect.\ref{mod-jet}). The CO
coma is modelled as the sum of an isotropic contribution and a
spiralling jet. The gas velocity and temperature are assumed to be
the same in the two components \citep[see discussion
in][]{boi+09}. Calculations performed with the second code are
presented in Sect.~\ref{sec:ir-jet}.

\subsection{Q-curve analysis}

To study the extent to which the spatial distribution of compounds
deviates from that expected for direct release from the nucleus,
\citet{disanti01} developed the method of $Q$-curves. An apparent
production rate $Q_{app}$ is derived from the intensity
$F_{line}$($x$) measured at a specific location $x$ along the
slit, under the assumption of spherical outflow from the nucleus
at constant velocity :

\begin{equation}
\label{eq:q-curve} Q_{app}(x) = \frac{4 \pi \beta \Delta^2
F_{line}(x)}{hc \nu g_{line} f(x)},
\end{equation}

\noindent where $g_{line}$ is the line fluorescence g-factor, and
$\beta$ is the CO photodissociation rate. The quantity $f(x)$
represents the fraction of the total number of molecules in the
coma within the region sampled at offset $x$. $Q$-curves trace the
evolution of $Q_{app}(x)$ as a function of $x$ (given in arc sec
or in km with respect to the nucleus position). For direct release
from the nucleus and optically thin conditions, $Q$-curves reach
quickly a terminal value equal to the true molecular production
rate. $Q_{app}$ values measured at small offsets from the nucleus
underestimate the true production rate as the measured intensities
are here affected by slit losses due primarily to seeing
\citep{disanti01}. The ratio between the terminal and on-nucleus
$Q_{app}$ values is larger for extended brightness distributions
than for more compact distributions. An extended distribution can
be identified by comparing with $Q$-curves obtained for dust or
species of purely nuclear origin observed in similar seeing
conditions. To examine how optical depth effects may have affected
the CO $Q$-curves, it is then important to take into account slit
losses in our modelling.

\subsection{Results with isotropic distributions}
\label{sec:Qcurve-iso}

\begin{figure}
\begin{center}
\includegraphics[angle=270, width=10cm]{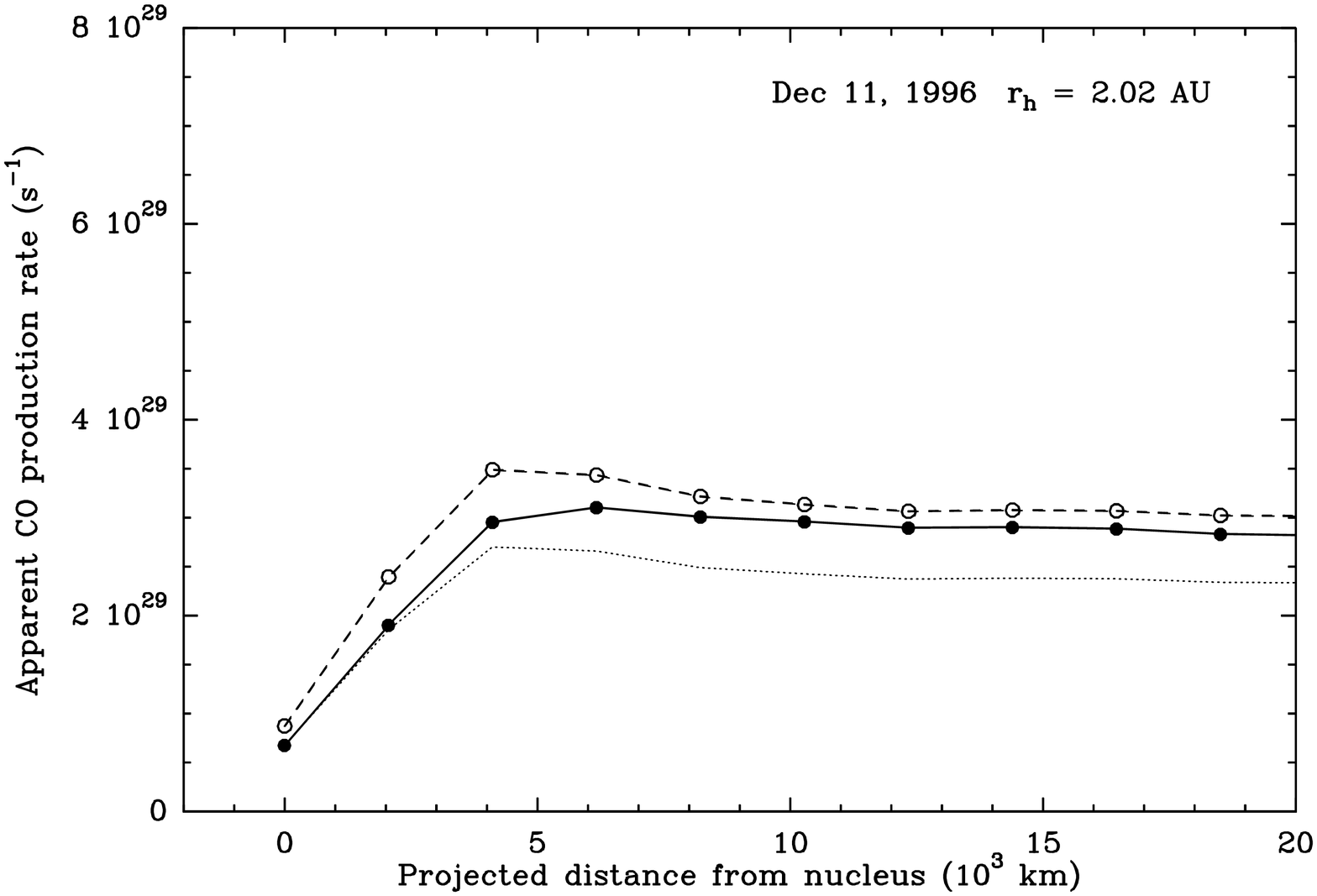}
\includegraphics[angle=270, width=10cm]{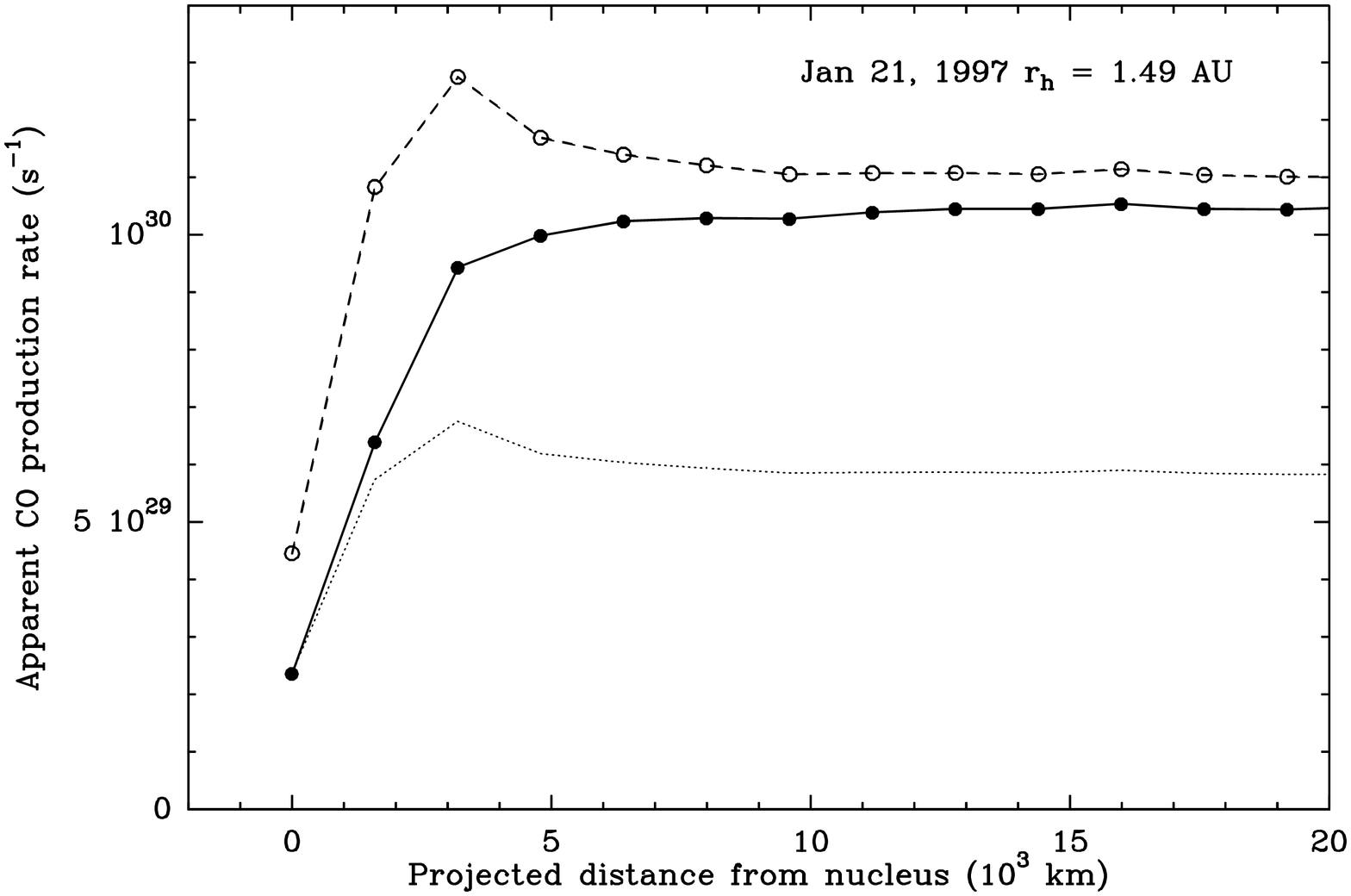}
\end{center}
 \caption{$Q$-curves generated with
the isotropic model of CO $v$ = 1--0 line emission. CO is assumed
to be released from the nucleus. Lines used are R2, R3, P2 and P3
(Dec. data), R0, R1, P1, R2, P2, R3, P3, R4, R5, R6 (Jan. data).
The aperture is set to 1\arcsec$\times$ 1\arcsec. These $Q$-curves
can be directly compared to Fig. 7 of \citet{disanti03}. Plain
symbols: results with all processes included (optical depth
effects, full rotational and vibrational excitation model). Open
symbols: reference $Q$-curve, i.e., model results under optically
thin conditions, assuming Boltzmann  rotational level populations
and ineffective vibrational collisional relaxation. Dotted curve:
optically thin $Q$-curve scaled to the value of the optically
thick $Q$-curve at the nucleus position. Calculations were
performed with $T$ = 60, 80 K, and $v_{exp}$ = 0.9, 1.0 km
s$^{-1}$, for December 11, 1996, and January 21, 1997 conditions,
respectively} \label{q-curve}
\end{figure}

In this section, we present results obtained under the assumption
of an isotropic CO coma. Figure~\ref{q-curve} shows $Q$-curves
computed for conditions corresponding to the data acquired on
December 11, 1996 ($r_h$ = 2.02 AU, $\Delta$= 2.83 AU, $Q_{\rm
CO}$ = 3.0 $\times$ 10$^{29}$ s$^{-1}$)  and January 21, 1997
($r_h$ = 1.49 AU, $\Delta$ = 2.2 AU, $Q_{\rm CO}$ = 1.1 $\times$
10$^{30}$ s$^{-1}$). Carbon monoxide is assumed to be solely
released from the nucleus. The CO ro-vibrational lines used for
generating these $Q$-curves are the same as those used by
\citet{disanti01} in their Fig.~7. According to \citet{disanti01},
the data of December 1996 are consistent with direct release of CO
from the nucleus, whereas those of January provide evidence for
production from an extended source.  We convolved the modelled
brightness distribution of CO $v$ = 1--0 emission lines by a 2-D
Gaussian representing the point spread function (PSF), and
computed expected fluxes in 1 $\times$ 1 \arcsec~boxes along the
radial direction, for direct comparison with the data of comet
Hale-Bopp presented by \citet{disanti01}. We assumed a seeing
equal to 3\arcsec~and 2.1\arcsec~for modelling the December and
January data, respectively. These values provide reference dust
$Q$-curves (see below) consistent with those observed. The gas
kinetic temperature was taken equal to 60 K (Dec. data) and 80 K
(Jan. data), based on the temperature measurements in the outer
coma from the CO IR lines \citep{disanti01}. Calculations shown in
Fig.~\ref{q-curve} were performed under the assumption that the
gas expansion velocity $v_{exp}$ is constant throughout the coma
and equal to 0.9 km s$^{-1}$ (Dec. data) and 1.0 km s$^{-1}$ (Jan.
data) \citep{biver02}.


For each date, two $Q$-curves were generated: 1) a reference
$Q$-curve (empty circles) where the lines were assumed to be
optically thin, collision quenching of fluorescence emission was
neglected, and the rotational level populations in the fundamental
were assumed to be at LTE; this reference $Q$-curve is also
representative of the expected $Q$-curve from dust, providing dust
emission falls off as 1/$\rho$, where $\rho$ is the projected
distance from the nucleus; 2) the second $Q$-curve (filled
circles) was generated from the intensities computed with the full
model. Apparent production rates were derived using
Eq.~\ref{eq:q-curve} and g-factors at $T$ = 60 K (Dec. data) and
80 K (Jan. data). The comparison of these $Q$-curves shows that
the CO ro-vibrational line emission was severely affected by
optical depth effects in January 1997 (Fig.~\ref{q-curve}). The
terminal $Q_{app}$ values for the scaled reference $Q$-curve and
CO $Q$-curve are in ratio $R_{Q}$ = 0.56. This is close to the
measured value of 0.49 \citep{disanti01}, showing that the opacity
of the lines explain in large part the extended nature of the
spatial brightness profiles. At the largest offsets plotted in the
figure, $Q_{app}$ is slightly below the true CO production rate
because opacity effects are still significant. As for the December
conditions, the model predicts that opacity effects should be
discernible on the $Q$-curve: the scaled reference $Q$-curve and
CO $Q$-curve are predicted to be in the ratio $R_{Q}$ = 0.83.
Figure~\ref{plot-opacity} plots the mean line opacity of several R
and P lines for the conditions of early March 1997 ($\Delta$ =
1.47 AU, $Q_{\rm CO}$ = 2 $\times$ 10$^{30}$ s$^{-1}$, $T$ = 90
K).  The mean line opacity is a weighted-average of the opacities
in spectral channels covering the Doppler-resolved line, where the
weight is the intensity in the spectral channels. The opacity of
the R6 line (4.9 for early March) drops to 2.2 and 0.5 for
mid-January and mid-December, respectively. Note that the IR CO
lines are generally much thicker than the radio lines
(Fig.~\ref{plot-opacity}).

As discussed by \citet{disanti01}, a reduction of the opacity of
the CO infrared lines may be expected if one considers gas
acceleration. To investigate this effect, we solved the radiation
transfer with gas accelerations consistent with the model of
\citet{combi99}. In the law used to interpret the December data,
the gas velocity increases from 0.81 to 0.92 km s$^{-1}$ for
radial distances between 1000 and 20\,000 km. As for January, it
increases from 0.81 (1000 km) to 1.03 km s$^{-1}$ (20\,000 km).
The results are shown in Fig.~\ref{q-curve2}. A small reduction of
the opacity effects is obtained. $R_{Q}$ values are increased from
0.83 to $\sim$0.90 and from 0.56 to 0.60, for December and
January, respectively. We consider that the December data are
satisfactorily explained by the model.

A 20\% discrepancy between the computed and measured $R_{Q}$
values remains for the January data, which indicates that the
observed spatial profiles are slightly more extended than expected
for a pure CO nuclear origin. Calculations of brightness profiles
applying to the February to May 1997 data of \citet{disanti01}
show that a discrepancy of 20--30\% is systematically present.
Figure \ref{q-curve-ext} shows $Q$-curves for January conditions
generated with combined 50:50 nuclear/extended distributions and
CO parent scale lengths $L_p$ from 1500 to 5000 km.  $R_{Q}$
decreases from 0.56 to 0.45 for $L_p$ increasing from 0 km (i.e.,
nuclear production) to 5000 km. Considering the reduction of
opacity effects caused by gas acceleration, the best fit to the
January data is obtained for $L_p = 5000$ km.



While we cannot exclude a contribution of a distributed source of
CO, its presence is not so compelling. Indeed, spatial profiles of
water obtained with the same high resolution infrared
instrumentation were also found to be broader than dust profiles
and profiles computed for a $\rho^{-1}$ brightness distribution
convolved with the PSF \citep{dello00}. The differences between
water and dust/PSF $Q$-curves are comparable to or larger than the
differences between observed and computed (pure nuclear origin) CO
profiles. \citet{dello00} discussed several possible explanations
for the different behaviors of water, dust and PSF $Q$-curves and
could not conclude for the contribution of a distributed source of
water. In the next section, we argue that the complexity of the CO
and dust Hale-Bopp comas should be considered to fully explain the
CO spatial profiles.


\subsection{Model with anisotropic CO distribution}
\label{sec:ir-jet}

As mentioned earlier, close to perihelion the coma of comet
Hale-Bopp was structured by a strong CO spiralling jet originating
from a low latitude region in the northern hemisphere
\citep{bock+09}. In contrast, the dust coma was dominated by a
high latitude northern jet producing repetitive shells
\citep{jorda99,vasun99}. This northern dust jet introduced a
significant departure between the nucleus position and the center
of brightness in visible images. Astrometric positions obtained
around perihelion were all off by $\sim$ --3\arcsec~in Declination
with respect to the nucleus position \citep{boi+07}. The
observations of the CO IR lines were performed with the slit
centred on the peak signal in the continuum and aligned along the
East-West direction \citep{disanti01,bro03}. Hence, they were
likely not centred on the nucleus position.


These spatial East-West profiles presented complex asymmetric
shapes, with the maximum of the emission frequently peaking at
1--2$ \times$ 10$^{3}$ km (typically 1--1.5\arcsec) West or East
of peak dust emission \citep{disanti01,bro03}. These complex
shapes might be related to the CO rotating jet.


We performed 3D radiative transfer calculations for CO infrared
lines using the coma model of \citet{bock+09}. We present results
obtained with their jet model (3) for the geometric conditions of
early March 1997, $Q_{\rm CO}$ = 2 $\times$ 10$^{30}$ s$^{-1}$ and
$T$ = 90 K. Several sets of IR CO data have been acquired during
this period \citep[namely on March 1.08, 3.80 and 5.99
UT][]{disanti01,bro03}, which is close to the date (11 March) of
the interferometric CO observations used to constrain this jet
model. The jet is defined by its half-aperture
($\Psi$~=~18.3\degre), latitude (\lat~=~20\degre), and the
fraction of CO molecules released within the jet (\fco~=~35.5\%).
In early March 1997, the spin axis had an aspect angle
$\theta_{\omega} \sim$ 77~\degre and a position angle on the sky
$pa_{\omega} \sim$ 205\degre based on the spin orientation derived
by \citet{jorda99}. Figure~\ref{plot-map} shows the expected
brightness distribution of the R6 CO line as a function of time
over one nucleus rotation period. The jet structure is aligned
along the RA axis towards West during about half of the period.
The brightness distribution is more extended along the jet. Using
the longitude of the jet at the nucleus surface determined by
\citet{bock+09} for 11 March UT and a synodic rotation period of
11.31$\pm$0.01 h \citep{farnham,jorda00}, we inferred that the
brightness distributions for 1.87, 3.80 and 5.99 March UT should
be approximatively those shown in panels 3, 4, and 11 of
Fig.~\ref{plot-map}, respectively. Next, we will also consider
panel 10 for comparisons with the 5.99 March UT data.

The 1.87 March UT spatial profile peaks East of the continuum peak
and falls off less rapidly with increasing $\rho$ towards West
than towards East \citep{disanti01}. This is consistent with
spatial profiles extracted from panel 3 (Fig.~\ref{plot-r6}). As
for the spatial profiles of 3.80 and 5.99 March UT, the sign of
the offset would match the data if the directions given in
\citet{bro03} were reversed. On 3.80 March the peak intensity is
observed westward the continuum peak and is more extended towards
East, while the model predicts the opposite (Fig.~\ref{plot-r6}).
This is puzzling because the time between the March 1.87 and 3.80
UT observations is 4.1 periods, so we expect the profiles to
exhibit the same shape. For 5.99 March UT, the peak intensity is
towards the East, but predicted towards West (panel 11). T. Brooke
(personal communication) comments that it is possible that the
orientations given in \citet{bro03} are in error though he does
not have firm evidence.

Figure~\ref{plot-r6} displays extractions parallel to the
East-West direction for several negative offsets $\delta$Dec in
Declination. The peak intensity decreases when the offset
increases, so that the spatial profile broadens. The position of
the centroid of the emission reaches 1 to 3.5\arcsec~for the times
corresponding to panels 3, 4 and 10. Very broad profiles are
obtained for panel 11 (expected to represent 5.99 March data),
because the jet is nearly in the plane of the sky and aligned
along East-West at that time (Fig.~\ref{plot-map}).

The intensity at 12\arcsec~normalized to the peak intensity is in
the range 0.12--0.20 at $\delta$Dec = 0 \arcsec, 0.13--0.25 at
$\delta$Dec=--1.2\arcsec, and 0.15--0.30 at
$\delta$Dec=--2.4\arcsec. Values as high as 0.3
($\delta$Dec=--1\arcsec) and 0.5 ($\delta$Dec=--2.5\arcsec) are
inferred in the West side of extractions from panel 2. For
comparison, on the spatial profiles recorded in March--April 1997,
this quantity generally ranges between 0.15--0.30, with extreme
values of 0.10 and 0.4 being also observed. Therefore, the
extended appearance of the CO brightness distribution could also
be partly related to the positioning of the slit. Detailed
modelling of the complex shapes of the CO IR radial profiles would
be challenging. They are strongly sensitive to the jet
characteristics and nucleus rotation properties. The
time-dependent model of \citet{bock+09} provides a very simplistic
description of Hale-Bopp's rotating CO coma, and does not explain
in detail the interferometric CO maps. More sophisticated models
would be required, and this is beyond the scope of this paper.

So far, we did not discuss how the absolute intensities given by
the model compare to the measurements. This exercise is possible
for the data published by \citet{bro03}, as intensities are listed
in the paper. Table~\ref{tab-intIR} lists measured R2 and R6 line
fluxes. Fluxes in a 2\arcsec$\times$1\arcsec~inner box centred on
the brightness peak ($F_{inner}$), and East-West averages of
5\arcsec$\times$1\arcsec~ boxes starting 3\arcsec~from the flux
peak ($F_{outer}$) are given. Results from model calculations are
also given in Table~\ref{tab-intIR} for three slit offsets in
Declination ($\delta$Dec = 0, --1.2, and --2.4\arcsec). To compare
to the 5.99 March data, we show results obtained using the radial
profiles from panel 10 instead of those from panel 11, as the
latter are much broader than those observed. The measured line
flux ratios $F_{outer}$/$F_{inner}$ ($\sim$ 1.04--1.09,
Table~\ref{tab-intIR}) can be explained providing the slit was
offset by 1--2\arcsec~ in Dec from the nucleus position. The
intensities measured on 3 March suggest that the CO production
rate was higher than 2 $\times$ 10$^{30}$ s$^{-1}$ at that time.
Using simple scaling, we infer a value consistent with the
\citet{bro03} determination of 2.6 $\times$ 10$^{30}$ s$^{-1}$
based on the line fluxes at large projected distances (outer box).
The intensities measured on 5 March suggest a total CO production
rate of $\sim$ 1.4 $\times$ 10$^{30}$ s$^{-1}$, consistent with
the value derived by \citet{bro03}. Note that the IR H$_2$O
observations performed on 5 March also suggest that comet
Hale-Bopp was less active on that day \citep{bro03}.

 Table~\ref{tab-intIR} provides also the intensities of the R2
and R6 lines obtained using a steady state isotropic model for the
CO density distribution. The flux ratios $F_{outer}$/$F_{inner}$
are, for most entries, slightly lower than the values obtained
with the anisotropic model (i.e., the model with the jet yields
East-West averaged radial profiles only slightly broader than for
an isotropic coma). We computed that, for a seeing of 1.7, the
expected $F_{outer}$/$F_{inner}$ ratio for isotropic and optically
thin conditions is 0.33, consistent with values measured for
CH$_4$ and H$_2$O in early March for similar seeing values
\citep{bro03}. Hence, the broad spatial distribution of the CO IR
line emission results mainly from optical depth effects and, at a
lesser extent, from the anisotropy of the CO coma and non-ideal
slit positioning.

 Our conclusion is inconsistent with the one obtained by
\citet{bro03} using similar modelling, but isotropic CO
distributions. \citet{bro03} analysed the radial profile of the R6
line observed on 5 March, and were able to fit the flux in the
inner box $F_{inner}$ with $Q$(CO) = 4 $\times$ 10$^{29}$ s$^{-1}$
and $T$ = 100 K. With such a low CO production rate, spatial
profiles were only slightly broader than in the optically thin
case. Using the same parameters, our model yields a value for
$F_{inner}$ which is 40\% lower than the observed value, and a
$F_{outer}$/$F_{inner}$ ratio equal to 0.43, i.e., barely higher
than the optically thin value. Our model requires a higher CO
production rate to fit the observations, yielding more pronounced
opacity effects. We were not able to elucidate the origin of the
discrepancy between the two models.

\subsection{Relative intensities of ro-vibrational lines}

We now examine how optical depth effects affect the relative
intensities of the CO $v$=1--0 ro-vibrational lines.

\begin{figure}
{\includegraphics[angle=270, width= \textwidth]{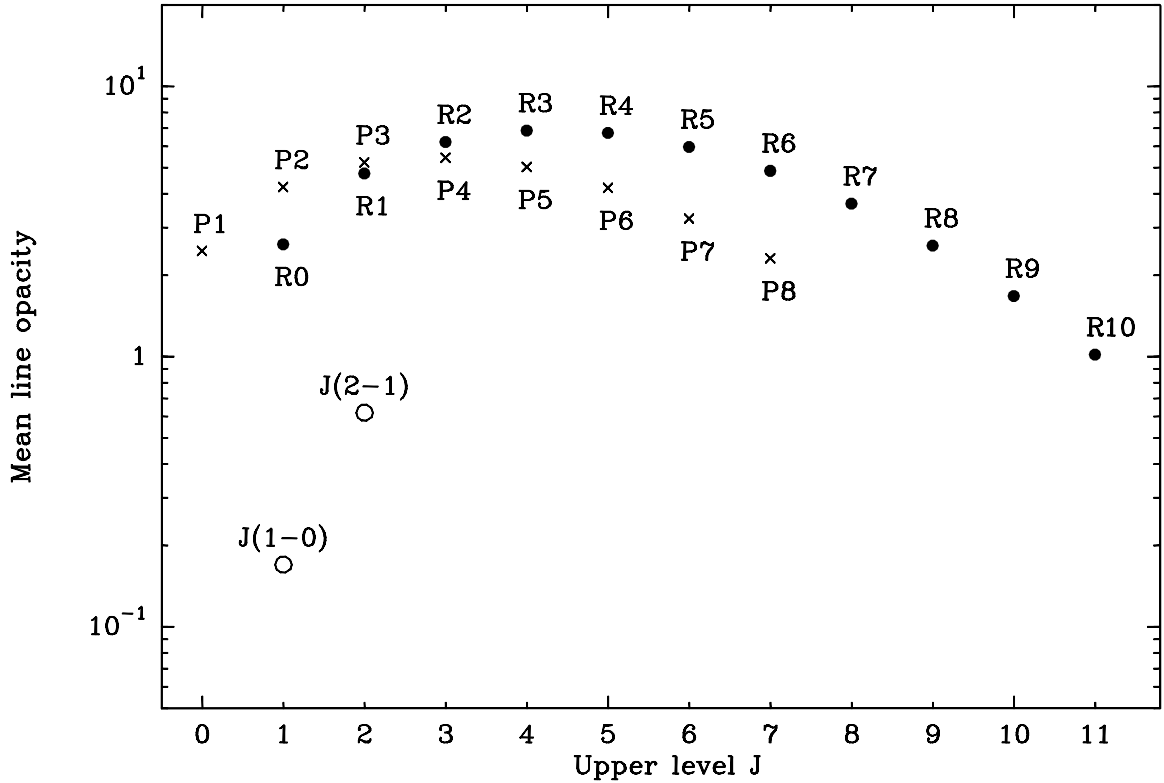}}
\caption{Mean opacity of observed CO IR lines for conditions of
early-March 1997 ($\Delta$ = 1.47 AU, $Q_{\rm CO}$ = 2 $\times$
10$^{30}$ s$^{-1}$, $T$ = 90 K, $v_{exp}$ = 1.1 km s$^{-1}$).
Calculations are for a circular aperture of 1\arcsec~diameter
centred on the nucleus. The mean opacities of the 115 and 230 GHz
lines in the same FOV are plotted, for comparison.}
\label{plot-opacity}
\end{figure}

\begin{figure}
\begin{center}
\includegraphics[angle=270,width=10cm]{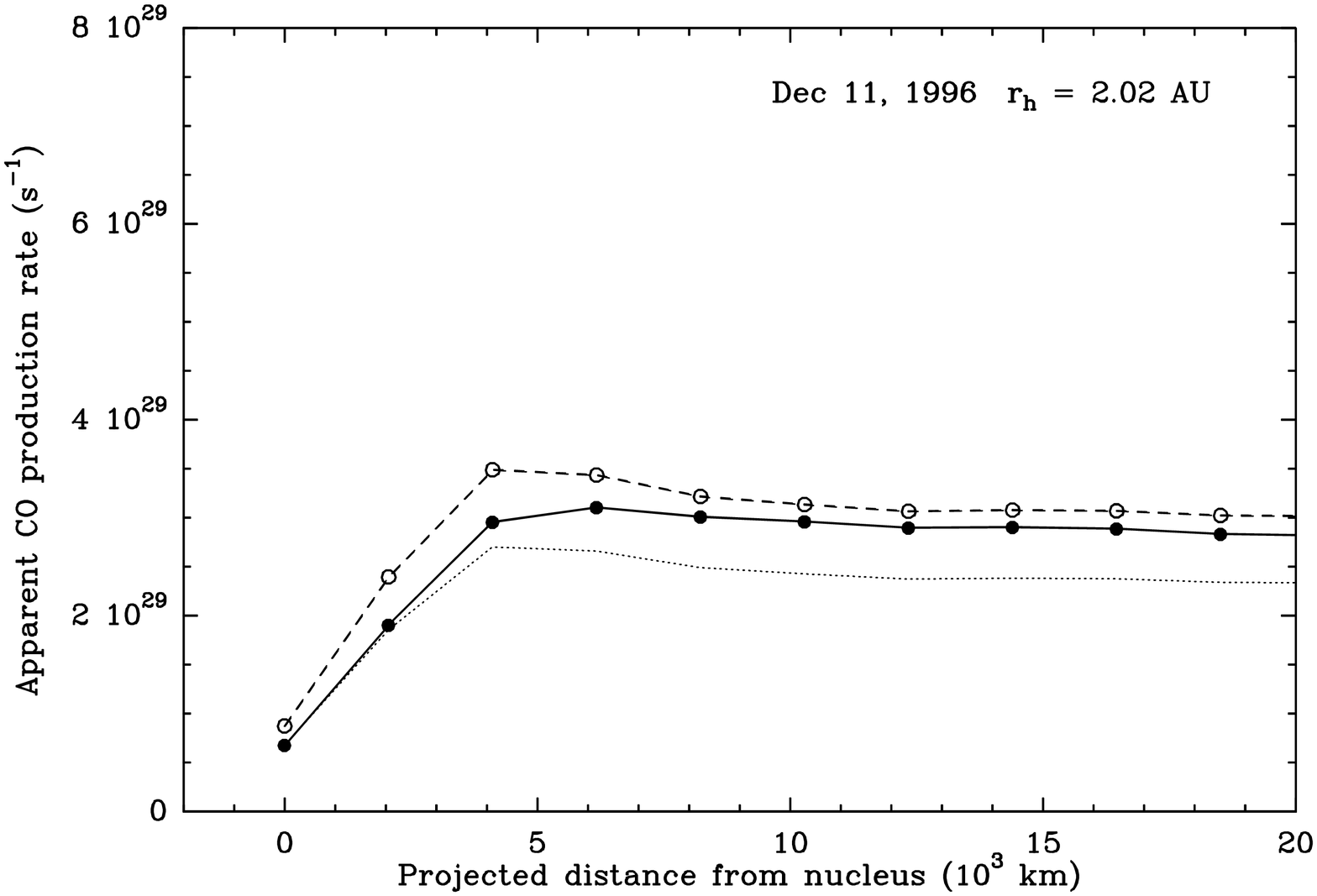}
\includegraphics[angle=270,width=10cm]{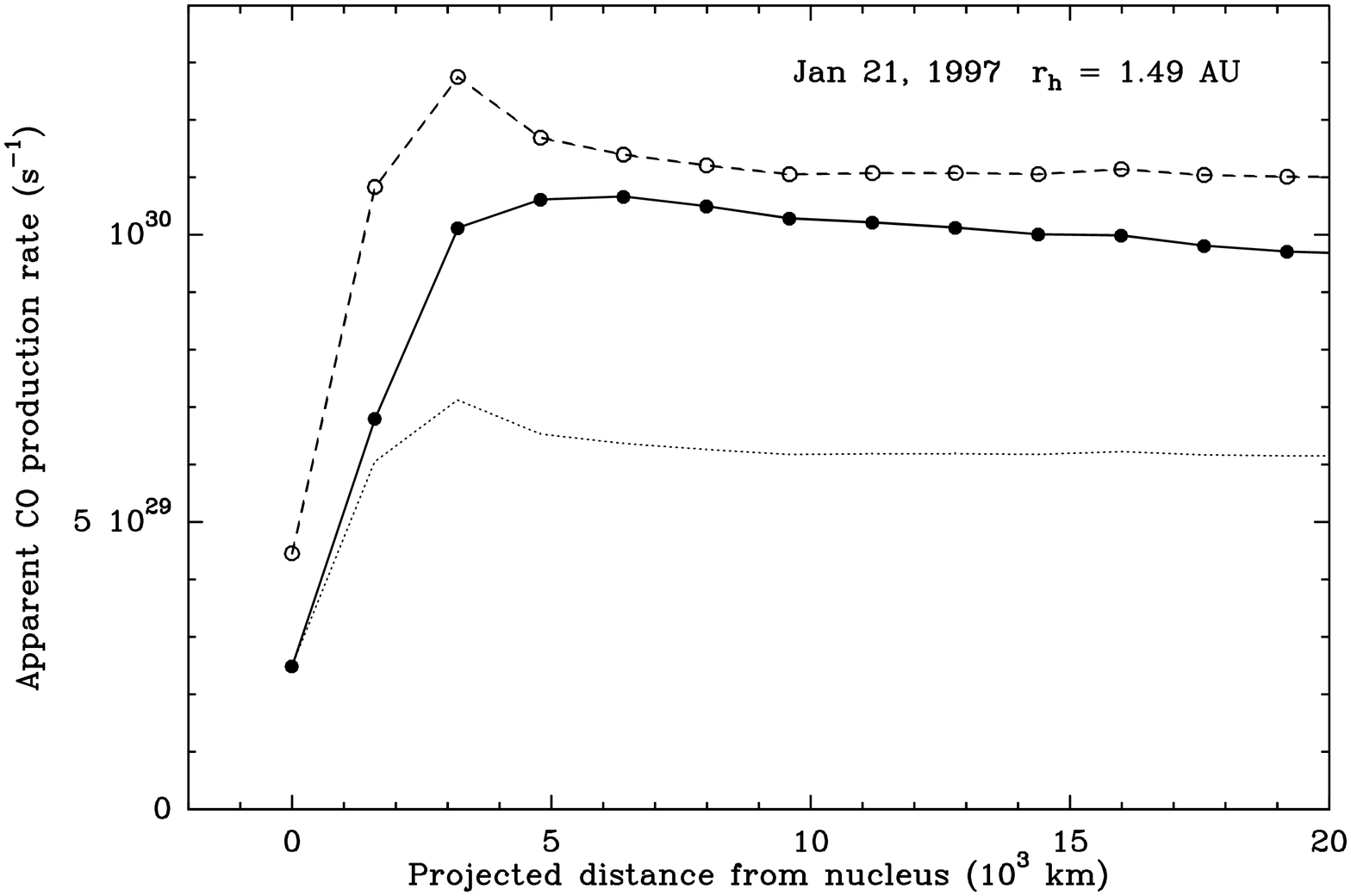}
\end{center}
\caption{Same as Fig.~\ref{q-curve} with gas acceleration
considered in radiation transfer calculations. Apparent CO
production rates were computed assuming a constant outflow
velocity $v_{exp}$ = 0.9 and 1.0 km s$^{-1}$, for 11 December
1996, and 21 January 1997, respectively.} \label{q-curve2}
\end{figure}

\begin{figure}
\resizebox{\hsize}{!}
{\includegraphics[angle=270,width=13cm]{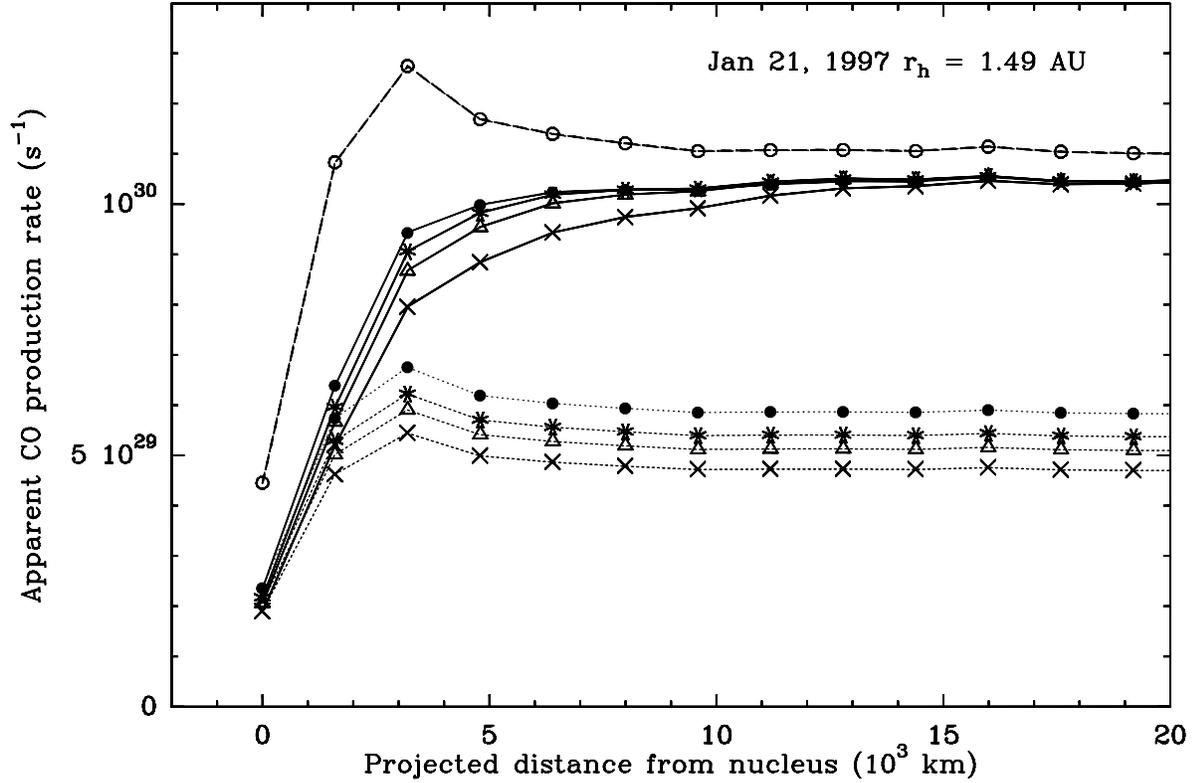}}
\caption{Same as Fig.~\ref{q-curve} for January 21, 1997,
considering both pure nuclear production and combined (50:50)
nuclear/extended distributions: nuclear (filled dots), combined
with $L_p$ = 1500 km (stars), 2500 km (empty triangles) and 5000
km (crosses). The reference $Q$-curve is shown by empty circles
connected by dashed lines, as in Fig.~\ref{q-curve}. Scaled
reference $Q$-curves are shown with dotted lines. Other model
parameters are the same as for the calculations shown in
Fig.~\ref{q-curve}.} \label{q-curve-ext}
\end{figure}

\begin{figure*}
\includegraphics[angle=270,width=16cm,bb = 230 50 550 780]{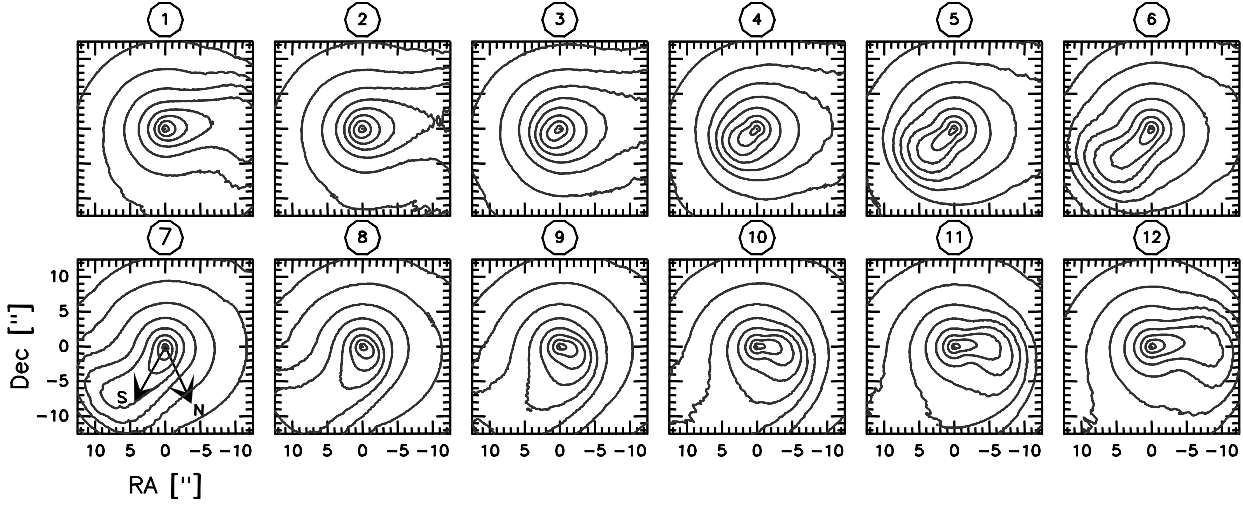}
 \caption{Time evolution of the brightness distribution (not corrected for seeing) of the $v$ = 1--0 R6 CO line for geometrical conditions of 1--5 March
 1997. The coma model of \citet{bock+09} is used (jet model (3); $Q_{\rm CO}$ = 2 $\times$ 10$^{30}$ s$^{-1}$). The gas
 coma temperature is taken equal to 90 K.
 Panels 1 to 12 cover one rotation period $P$ = 11.35 h with a time step of
 $P$/12. Line fluxes were computed
 for 0.2$\times$0.2\arcsec~pixel size. Contour levels are in
 logarithmic scale and spaced by 0.15.  Maximum fluxes on the maps are between 9.2 and
 9.6 10$^{-18}$ W m$^{-2}$. The nucleus North pole (N) and the Sun (S)
 directions are indicated. Celestial East is on the left.} \label{plot-map}
\end{figure*}

\begin{figure}
\begin{minipage}{9cm}
\includegraphics[angle=270,width=8cm]{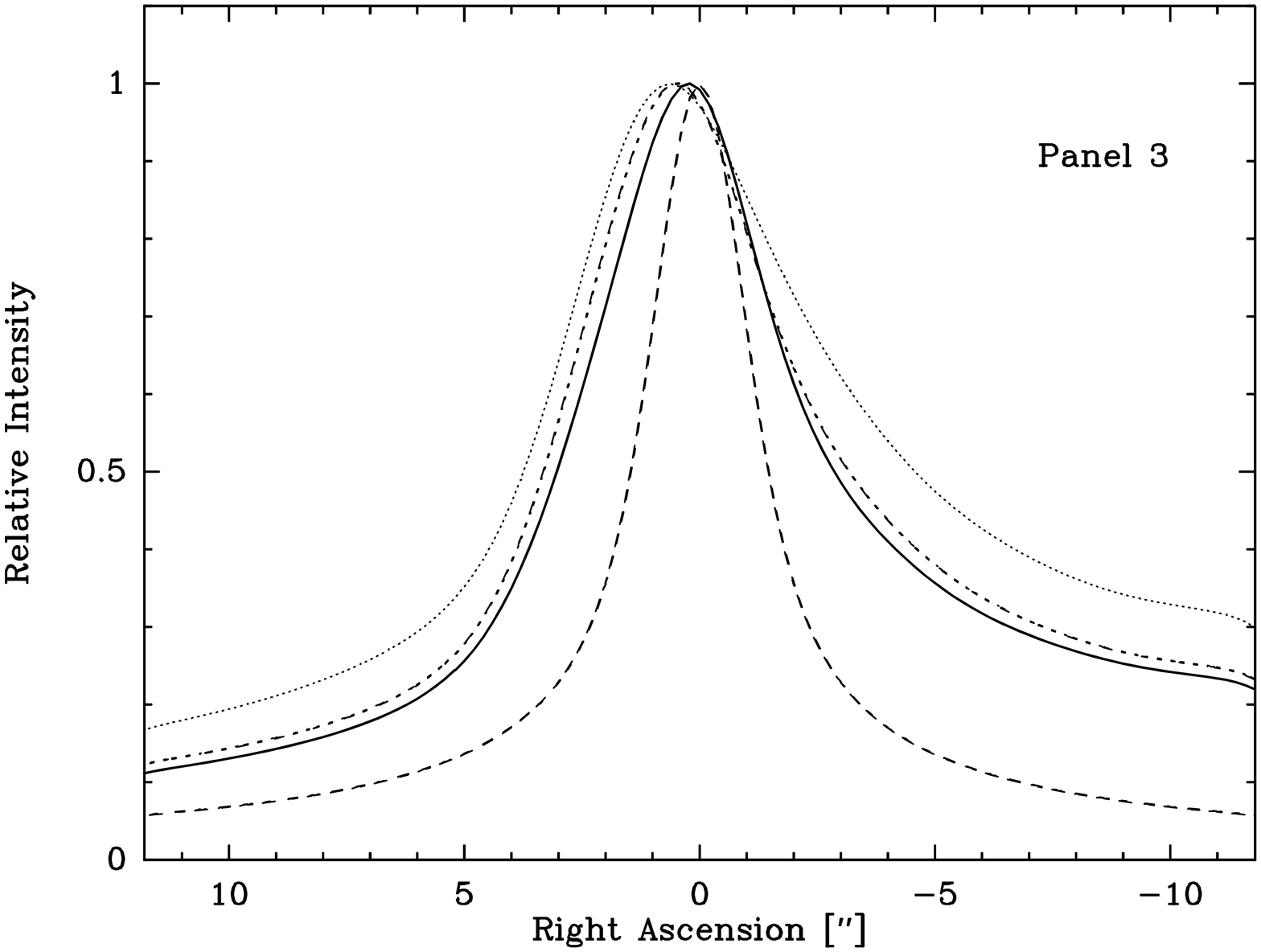}
\end{minipage} \hfill
\begin{minipage}{9cm}
\includegraphics[angle=270,width=8cm]{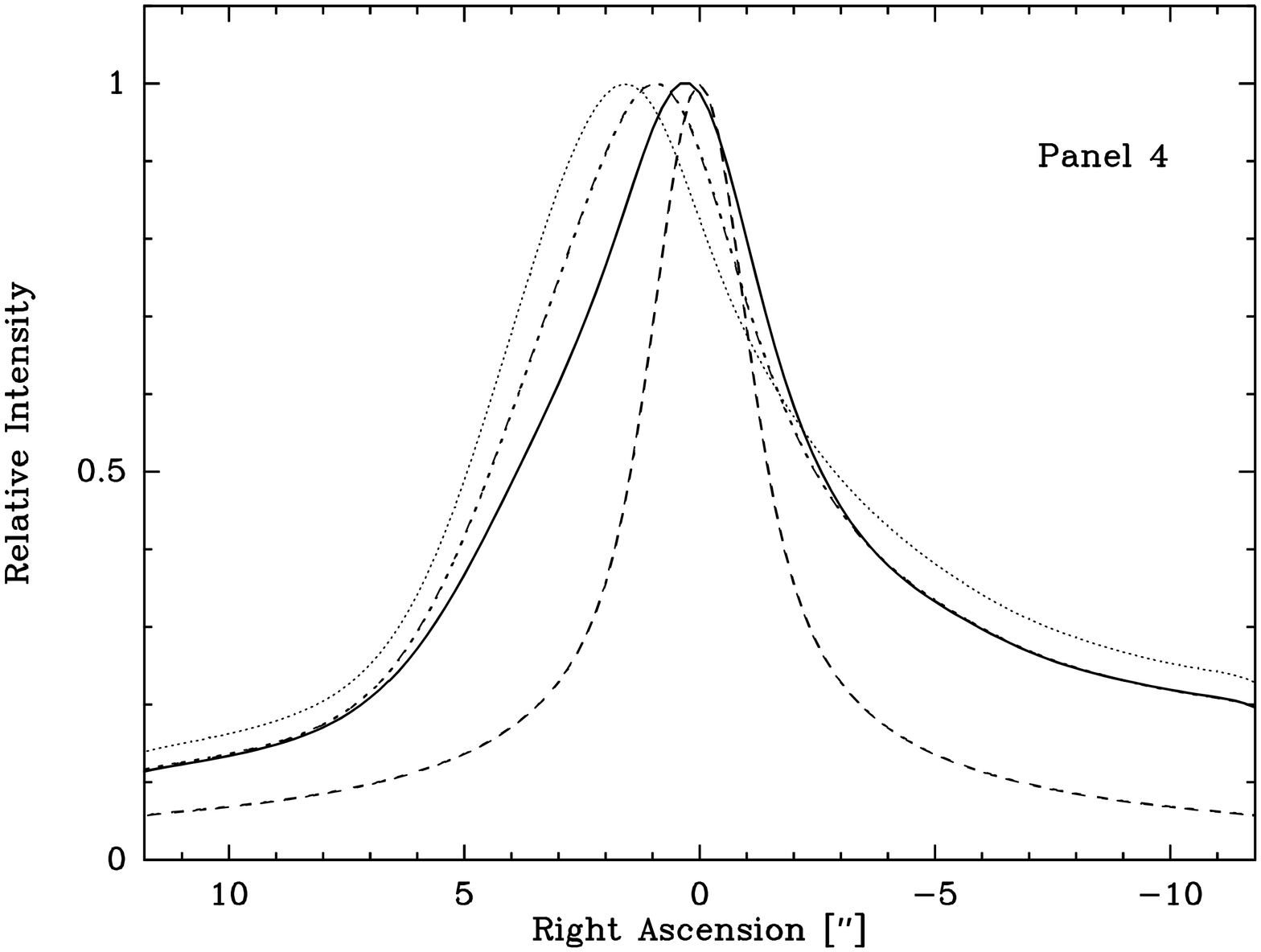}
\end{minipage}
\begin{minipage}{9cm}
\includegraphics[angle=270,width=8cm]{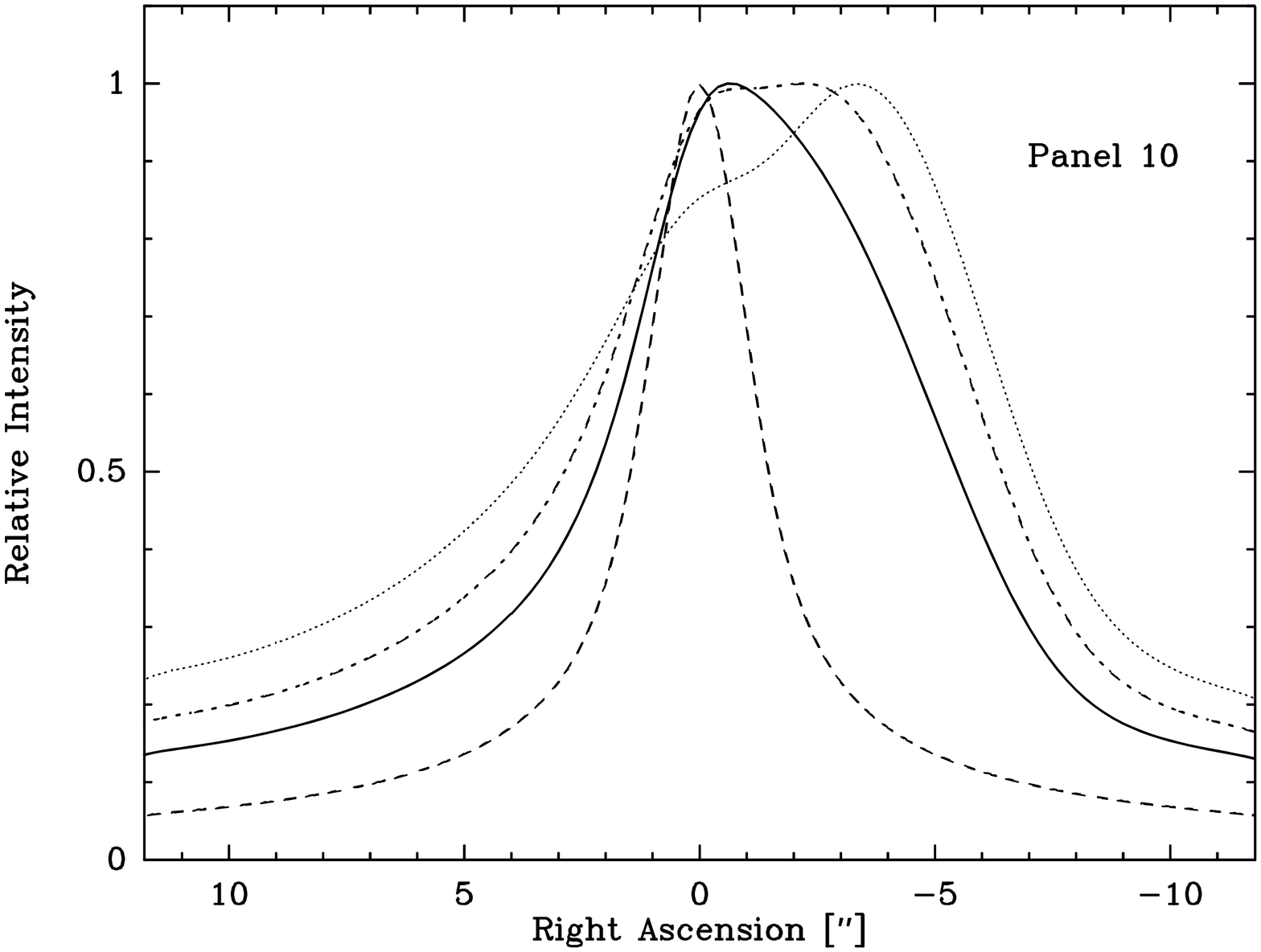}
\end{minipage} \hfill
\begin{minipage}{9cm}
\includegraphics[angle=270,width=8cm]{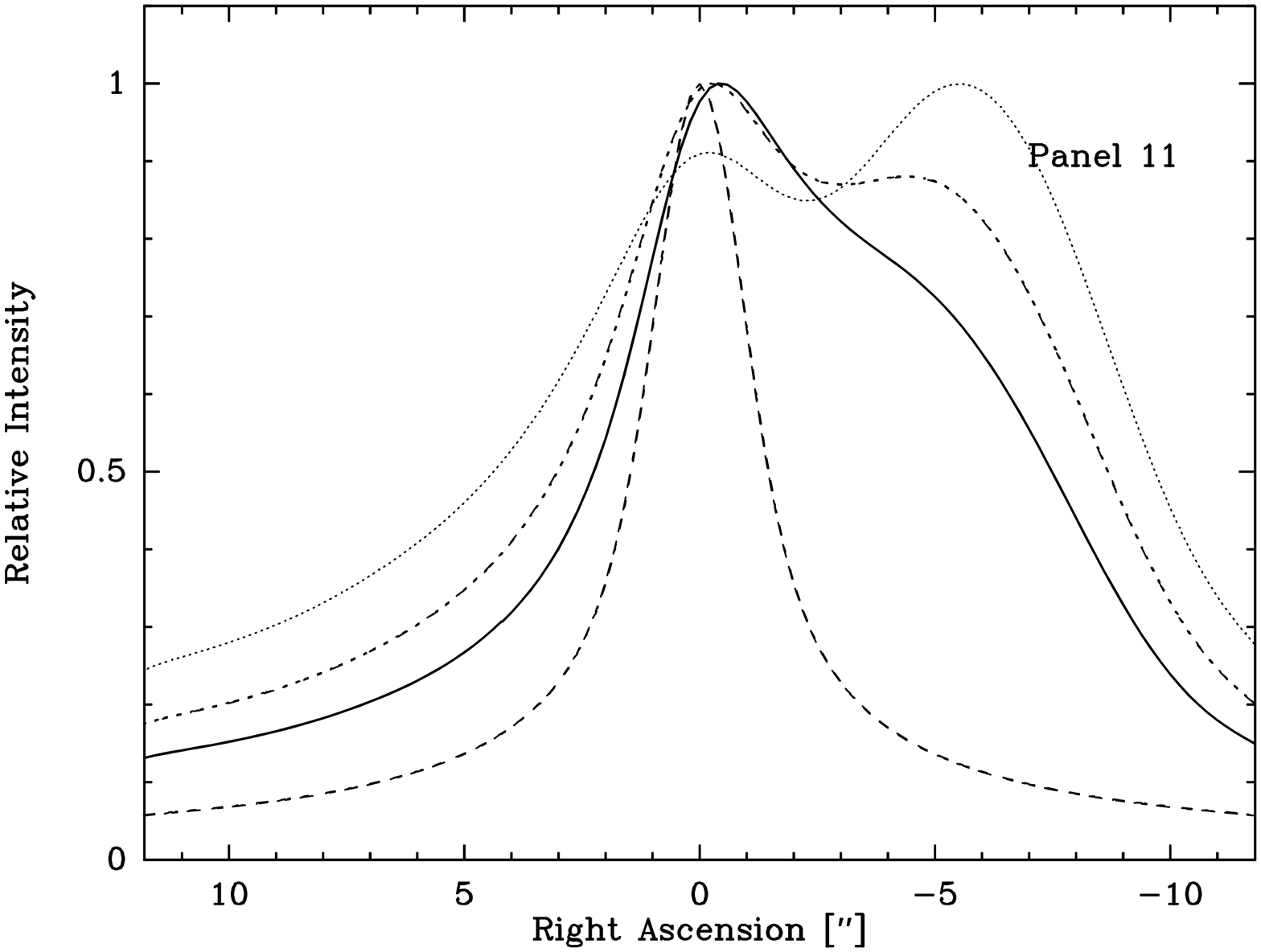}
\end{minipage}
 \caption{CO $v$ = 1--0 R6 spatial profiles derived from the brightness
 maps (Panels 3, 4, 10, and 11) shown
 in Fig.~\ref{plot-map} convolved with the PSF. Extractions are
parallel to the RA axis and centred
 at $\delta$Dec = 0\arcsec~(plain line), $\delta$Dec = --1.2\arcsec~(dashed-dotted line) and  $\delta$Dec = --2.5\arcsec~(dotted line).
 Intensity has been integrated
 over rectangular boxes of 1\arcsec~along Dec axis and 0.2\arcsec~along RA
 axis. The dashed line shows a $\rho^{-1}$ column density profile convolved with the PSF. The seeing
was taken equal to 1.7 \arcsec. Peak intensities (units of
10$^{-17}$ W m$^{-2}$) are 3.5, 3.2, 2.4 (Panel 3), 3.5, 3.4, 2.8
(Panel 4), 3.4, 2.7, 2.1 (Panel 10), and 3.3, 2.6, 1.9 (Panel 11)
for $\delta$Dec = 0, --1.2\arcsec, and --2.4\arcsec,
respectively.} \label{plot-r6}
\end{figure}

\begin{figure}
\begin{center}
\includegraphics[angle=270,width=13cm]{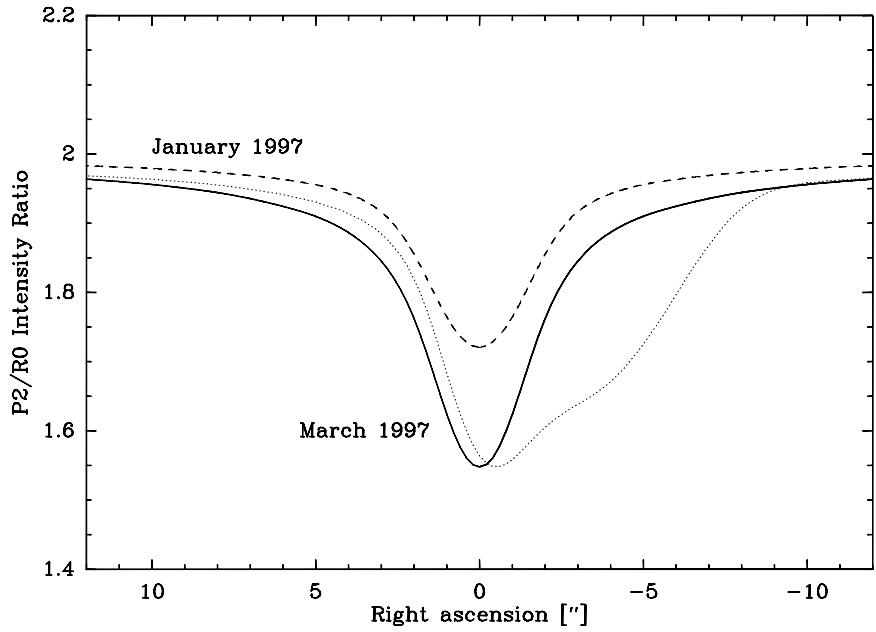}
\end{center}
 \caption{Ratio of the P2 to R0 line intensities along a slit aligned along the RA axis and centred on the nucleus. Results from
 radiative transfer calculations in an isotropic coma are shown in dashed (parameters of 21 Jan.
1997) and plain (early March 1997) lines. Results obtained with
the jet model (phase of panel 10) are shown in dotted lines. The
ratio was computed using PSF-corrected intensities over
rectangular boxes of 1\arcsec~along Dec axis and 0.2 \arcsec~along
RA
 axis.} \label{plot-p2tor0}
\end{figure}

As already mentioned, for beams passing close to the nucleus, the
ratio of the P2 to R0 line intensities was measured to be lower
than the statistical value of 2 expected under optically thin
conditions \citep{disanti01}. The East-West averaged P2/R0
intensity ratio was found to approach the statistical ratio at
distances $\rho$ $>$ 2300 km ($>$ 1.5\arcsec) in January, and
$\rho$ $>$ 5800 km ($>$ 5.5\arcsec) in March--April. This trend
can only be explained by opacity effects. As shown in
Fig.~\ref{plot-opacity}, at the temperatures encountered in the
Hale-Bopp coma, the opacity of the R0 line is smaller than the P2
line opacity (by a factor of 1.6 for January--April). Figure
\ref{plot-p2tor0} plots the P2/R0 intensity ratio derived from our
radiative transfer models for the two periods. The on-nucleus
value of 1.55 found for early March agrees with (although somewhat
higher) the measured value of $\sim$1.3$\pm$ 0.1 derived from Fig.
A1 of \citet{disanti01}, not considering the possible slit offset
in Dec discussed in the previous section (in which case we may
conclude that the model slightly underestimate optical depth
effects). The computed evolution of the P2/R0 intensity ratio
along the slit agrees with the measurements. As for January,
radiative transfer calculations in the assumption of an isotropic
coma lead to a P2/R0 ratio lower than the observed value ($\sim$
1.3$\pm$0.2). Our modelling does not predict a significant
decrease of the P3/R1 intensity ratio with respect the optically
thin statistical value of 1.5, because the two lines have
approximately the same opacity (Fig.~\ref{plot-opacity}). A P3/R1
intensity ratio significantly lower than 1.5 was observed,
however, but only in January \citep[Fig. A1 of][]{disanti01}.
Overall, there are some indications that our modelling
underestimates opacity effects, which strengthens our
interpretation of the CO spatial profiles as being strongly
affected by opacity effects.

In optical thin conditions, the rotational temperature
$T_{rot}^{v}$ describing the population distribution in the
vibrational excited state can be retrieved from the relative
intensities of the ro-vibrational IR lines. This temperature is
close to the rotational temperature in the ground vibrational
state $T_{rot}$, and probes the gas kinetic temperature when the
field of view encompasses essentially the collision-dominated
coma. Applying the method of rotation diagrams to the CO Hale-Bopp
data, \citet{disanti01} retrieved $T_{rot}^{v}$ for $v$=1, as a
function of projected distance to nucleus. However, this method is
in principle not appropriate since the CO lines are optically
thick. We applied the same method to synthetic line fluxes, giving
the same weight to all lines. We found that : i)
$T_{rot}^{v}(meas.)$ $\leq$ $T_{rot}^{v}$, where
$T_{rot}^{v}(meas.)$ is the value retrieved from the rotation
diagram; ii) $T_{rot}^{v}(meas.)$ and $T_{rot}^{v}$ differ by an
amount which depends on the lines used in the rotation diagram;
iii) $T_{rot}^{v}$ is larger than $T_{rot}$ in the ground
vibrational state by typically 8--10\%. For the December
simulations performed with $T$ = 60 K (R2--R3, P2--P3 lines), we
found for the inner 1\arcsec$\times$1\arcsec~box:
$T_{rot}^{v}(meas.)$ = 60 K, $T_{rot}^{v}$ = 66 K; for the January
simulations with $T$ = 80 K (R0 to R6, P1 to P4 lines):
$T_{rot}^{v}(meas.)$ = 65 K, $T_{rot}^{v}$ = 86 K. The rotation
diagram method underestimates more significantly $T_{rot}^{v}$
when the R0, R1 and P1 to P3 lines are considered, because these
lines, which are pumped from the lowest energy $v$ = 0 rotational
states, are less optically thick than the other lines
(Fig.~\ref{plot-opacity}). \citet{disanti01} points out that the
measured temperatures on 24 February 1997 are higher compared with
those on most other dates. We suggest that this is because the P1
to P3 and R0--R1 lines were not used (they were not observed on 24
February). In summary, $T_{rot}^{v}$ values derived from set of
lines that include R0--R1 and P1 to P3 lines likely underestimate
(by typically 10--15 K) the gas kinetic temperature $T$ in the
innermost lines of sight.

\section{Summary}
\label{sec:summary}

The high productivity of comet Hale-Bopp in carbon monoxide made
possible the investigation of the spatial distribution of this
molecule in the inner ($r$ = 1000--10\,000 km) coma by means of
spectroscopy from ground-based telescopes. Because there was
observational evidence of an extended production of CO in infrared
data \citep{disanti01,bro03}, we performed a detailed analysis of
the CO data acquired with the IRAM Plateau de Bure interferometer
close to perihelion (March 1997) for a confirmation of this
extended production. This study benefited from the simultaneous
observations of two CO rotational lines, the combination of both
high and low angular resolutions probing coma radii from $\sim$800
to $\sim$20\,000 km, and last but not least very high
signal-to-noise ratios. The results obtained from this study urged
us to analyze the CO infrared data using radiative transfer codes
similar to those used for interpreting the millimeter data. We
reached the following conclusions:

\begin{enumerate}

\item Both $J$(1--0) and $J$(2--1) radio lines exclude pure CO
production from an extended source, unless the characteristic
parent scale length $L_{\rm p}$ of the distributed source was
 smaller than 500 km;

\item  Composite models considering half of the CO production from
a distributed source cannot explain the radio observations, unless
$L_{\rm p}$ was $<$ 1500 km. A good fit to the data is obtained
when CO is assumed to be solely released from the nucleus;

\item The $J$(1--0) line data are best explained when considering
the radial evolution of the temperature in the coma due to
photolytic heating. Secure results are obtained for the $J$(2--1)
line thanks to the development of a radiative transfer code which
accounts for optical depth effects in the emergent emission;

\item The presence of a CO spiralling jet in the Hale-Bopp's coma
does not affect much our analysis because the millimetric
observations spanned a large fraction of the nucleus rotation and
probed almost the whole coma at projected distances $\rho < $
20\,000 km;

\item The observed CO ro-vibrational lines are optically thick.
The broad extent of the CO brightness distribution in the infrared
is mainly due to optical depth effects entering in the emitted
radiation. Surprisingly, our conclusion is not consistent with the
one obtained by \citet{bro03} who performed similar radiative
transfer calculations. Comparison of model results shows that they
quantitatively differ.

\item The slit-oriented spatial profiles of the CO infrared lines
are strongly affected by the CO jet, making their interpretation
difficult. They are more extended than expected for a pure nuclear
origin possibly also because the slit was centred on the peak of
the dust continuum emission which departed from the nucleus
position.

\item The comparison between observed and computed P2/R0 line
intensity ratios suggests that our modelling underestimates
optical depth effects, which strengthens our interpretation of the
broad extent of the IR emission.

\item When optical depth effects are present, the rotation diagram
method introduces errors in rotational temperature determinations.

\end{enumerate}

In conclusion, our analysis shows that there is no compelling
evidence of extended production of CO in Hale-Bopp's coma. We
anticipate that a similar conclusion should be reached for comet
Hyakutake from a reanalysis of the observed IR lines.

\acknowledgements {\bf Acknowledgements:} IRAM is an international
institute co-funded by the CNRS, France, the
Max-Planck-Gesellschaft, Germany, and the Instituto Geogr\`afico
Nacional, Spain. This work has been supported by the Programme
national de plan\'etologie of Institut national des sciences de
l'univers.

\clearpage

\appendix
\begin{center}
 {\bf APPENDIX : Visibilities for long-lived parent molecules}
\end{center}

In this Appendix, we demonstrate that, for a parent molecule
distribution, the amplitude of the visibility
$\bar{\mathcal{V}}(\sigma)$ varies with the $uv$-radius $\sigma$
according to $\sigma^{-1}$, in first approximation.


Let us start from the general equation of the complex visibility
$\mathcal{V}(u,v)$ that can be approximated to:


\begin{eqnarray}
\label{app-eq1} \mathcal{V}(u,v) & = & \frac{c}{\nu}
\int_{-\infty}^{+\infty} \int_{-\infty}^{+\infty}
A(x,y) F(x,y) \nonumber \\
 & & \times \ \exp(-\frac{2i\pi \nu}{c} (ux+vy))\, dx dy
\end{eqnarray}

\noindent in units of line area (Jy km s$^{-1}$). $u$ and $v$ are
the coordinates of the baseline vector for two antennas in the
$uv$-plane. $F(x,y)$ is the brightness distribution in the plane
of the sky. $A(x,y)$ is the antenna primary beam pattern of
half-power beam width $\theta_{\rm B}$, approximated by a 2-D
Gaussian centred on $(0,0)$ given by:

\begin{eqnarray}
A(x,y) =  \exp(- \frac{4\ln(2)(x^2+y^2)}{\theta_{\rm B}^2}).
\nonumber\end{eqnarray}

Under optically thin conditions and assuming that the population
of the upper level of the transition $p_u$ does not vary with
distance to nucleus, $F(x,y)$ is proportional to
$\int_{-\infty}^{+\infty} n(r) dz$, where $n(r)$ is the local
density of the considered molecule at distance $r$ from the
nucleus. If we assume a Haser parent molecule distribution and
neglect the photodestruction of the molecule, then:

\begin{eqnarray}
F(x,y) & = & K \frac{1}{\sqrt{x^2+y^2}} = \frac{K}{\rho},\nonumber
\end{eqnarray}

\noindent $\rho$ is in radians. $K$ is a constant factor equal to:

\begin{eqnarray}
K = \frac{ h \nu A_{ul} p_u Q}{16 \pi v_{\rm exp} \Delta}.
\nonumber\end{eqnarray}

$Q$ is the gas production rate and $v_{\rm exp}$ is the gas
expansion velocity. $\Delta$ is the geocentric distance. In polar
coordinates in the Fourier ($u = \sigma\cos{\alpha}$, $v =
\sigma\sin{\alpha}$) and image ($x = \rho\cos{\theta}$, $y =
\rho\sin{\theta}$) planes, Eq.(\ref{app-eq1}) reduces to:

\begin{eqnarray}
\label{app-eq2} \mathcal{V}(\sigma,\alpha)=\frac{c K}{\nu}
\hspace{-0.1cm}\int_{0}^{+\infty}\hspace{-0.3cm}\int_{0}^{2\pi}
\hspace{-0.4cm}A(\rho)
 \exp(-\frac{2i\pi \nu}{c} \rho \sigma\cos(\alpha-\theta))\, d\theta d\rho.
\nonumber \\
\phantom{...}
\end{eqnarray}

\noindent It can be demonstrated that:

\begin{eqnarray}
\int_{0}^{2\pi} \exp(-\frac{2i\pi \nu}{c} \rho \sigma
\cos(\alpha-\theta))\, d\theta = 2\pi {\rm J}_0(\frac{2\pi \nu
\rho \sigma}{c}), \nonumber\end{eqnarray}

\noindent where J$_0$ is the Bessel function of first kind and
zero order. Therefore, Eq.(\ref{app-eq2}) reduces to:

\begin{eqnarray}
\mathcal{V}(\sigma,\alpha) = \frac{2 \pi c
K}{\nu}\int_{0}^{+\infty} A(\rho) {\rm J}_0(\frac{2\pi \nu \rho
\sigma}{c}) d\rho. \nonumber
\end{eqnarray}

\noindent $\mathcal{V}(\sigma,\alpha)$ is a real function (a
general property of the Fourier Transform of a symmetric function)
of amplitude $\bar{\mathcal{V}}(\sigma)$ =
$\mathcal{V}(\sigma,\alpha)$, which can be also written:

\begin{eqnarray}
\label{app-eq3} \bar{\mathcal{V}}(\sigma) =  \frac{c^2 K}{\nu^2}
\times \frac{1}{\sigma}\int_{0}^{+\infty} A(\frac{c \rho}{ 2 \pi
\nu \sigma}) {\rm J}_0(\rho) d\rho. \nonumber
\end{eqnarray}

Let us call $D$ the diameter of the antennas. For $\sigma/D$ $\gg$
0.2 (typically $\sigma$ $>$ 20 m for 15-m dishes), the $A$ term in
the integral can be considered as unity. As $\int_{0}^{+\infty}
{\rm J}_0(\rho) d\rho$ = 1, then:

\begin{eqnarray}
\bar{\mathcal{V}}(\sigma) =  \frac{ h c^2 A_{ul} p_u Q}{16 \pi \nu
v_{\rm exp} \Delta} \times \frac{1}{\sigma}.
\end{eqnarray}

This formula remains valid when photodissociation occurs, provided
that $\sigma \gg c \Delta/\nu L$, where $L$ is the
photodissociation scale length.

In the same units (Jy km s$^{-1}$), the line area for
autocorrelation (ON--OFF) spectra which corresponds to
$\bar{\mathcal{V}}$(0) is:

\begin{eqnarray}
\bar{\mathcal{V}}(0) =  \frac{ h c A_{ul} p_u \cal{N}}{4 \pi
\Delta^2},
\end{eqnarray}

\noindent where $\cal N$ is the number of molecules sampled by the
primary beam, which can be computed by volume integration.

\clearpage
\begin{table}
\caption{Observed intensities of CO lines in single-dish (SD) and
interferometric (Int.) modes.} \label{tab-int}
\begin{center}
\begin{tabular}{lccc}
\hline \noalign{\smallskip} \multicolumn{1}{l}{Line} &
\multicolumn{1}{c}{Mode} &
\multicolumn{1}{c}{Beam} & \multicolumn{1}{c}{Intensity$^{\footnotesize a}$}  \\
& & & (Jy km s$^{-1}$) \\
\noalign{\smallskip} \hline \noalign{\smallskip}
$J$(1--0) & SD &41.8\arcsec &    10.8 $\pm$ 0.43   \\
         & Int. & 2.57\arcsec$\times$3.58\arcsec &   0.93 $\pm$ 0.04  \\
\hline \noalign{\smallskip}
& & $R_{\rm 1-0}$ = $F_{\rm SD}/F_{\rm Int}$ & 11.6 $\pm$ 1.8$^{\footnotesize b}$ \\
\hline \noalign{\smallskip}
$J$(2--1) & SD & 20.9\arcsec  &    82.8 $\pm $0.59    \\
         & Int.  & 2.0\arcsec$\times$1.38\arcsec   &   4.65 $\pm $0.15  \\
\hline \noalign{\smallskip}
& & $R_{\rm 2-1}$ = $F_{\rm SD}/F_{\rm Int}$ & 17.8 $\pm$ 3.8$^{\footnotesize b}$ \\
\hline
\end{tabular}
\end{center}

$^{\footnotesize a}$ From \citet{bock+09}.

$^{\footnotesize b}$ Include uncertainties in flux calibration:
10\% for CO $J$(1--0), 15\% for CO $J$(2--1) \citep{bock+09}.


\end{table}

\begin{landscape}

\begin{table}
 \caption{Modelled PdBI intensities of
the CO lines in comet Hale-Bopp and slope of the visibility
curves.} \label{tab-int-mod}
\begin{tabular}{lcccccccccc}
 \hline \noalign{\smallskip} \multicolumn{1}{l}{Model} &
\multicolumn{1}{c}{ $Q_{\rm CO }^N$:$Q_{\rm CO
}^E$$^{\footnotesize a}$ } &
\multicolumn{1}{c}{$T$} & \multicolumn{1}{c}{$v_{\rm exp}$} & \multicolumn{1}{c}{$L_p$} & \multicolumn{2}{c}{$F_{\rm Int}$} & \multicolumn{2}{c}{$R_{\rm mod}$}  & \multicolumn{2}{c}{Slope$^{\footnotesize b}$} \\
& (\%:\%) & (K) & (km s$^{-1}$) & (km) & \multicolumn{2}{c}{(Jy km s$^{-1}$)}  & & \\
\cline{6-11}\noalign{\smallskip}
& & & & & $J$(1--0) &  $J$(2--1) &$J$(1--0) &  $J$(2--1) &$J$(1--0) &  $J$(2--1) \\
\hline \noalign{\smallskip}
Nuclear+Thin$^{\footnotesize c}$ & 100:0 & 120 & 1.05 & 0 & 0.74 & 5.86 & 15.0 & 14.7 & -1.02 & -1.04 \\
Nuclear      & 100:0 & 120 & 1.05 & 0 & 0.70 & 4.25 & 15.7 & 19.1 & -1.07 & -1.20 \\
Nuclear      & 100:0 & $T_{var}$$^{\footnotesize d}$ & 1.05 & 0 & 1.00 & 4.64  & 11.6  & 18.7 & -0.96 & -1.42 \\
Nuclear      & 100:0 & 120 & $v_{\rm var}$$^{\footnotesize e}$ & 0 & 0.78  & 4.77  & 13.6 & 17.2 & -1.04 & -1.21 \\
Nuclear      & 100:0 &$T_{var}$$^{\footnotesize d}$ & $v_{\rm var}$$^{\footnotesize e}$ & 0 & 1.10 & 4.94 & 10.3 & 17.7 & -0.95 & -1.50  \\
Nuclear      & 100:0 &$T_{var(alt)}$$^{\footnotesize f}$ & $v_{\rm var}$$^{\footnotesize e}$ & 0 & 1.09 & 5.38 & 10.7 & 16.3 & -0.98 & -1.35  \\
Extended    & 0:100 & 120 & 1.05 & 2000 & 0.32  & 1.31 & 32.2 & 55.8 & -1.75 & -1.96 \\
Extended    &  0:100 & $T_{var}$$^{\footnotesize d}$& $v_{\rm var}$$^{\footnotesize e}$ & 2000 & 0.53 & 2.27 & 19.6 & 34.3 & -1.44 & -1.82   \\
Composite   & 50:50 & 120 & 1.05 & 2000 & 0.51 & 2.92  & 20.8  & 26.5 & -1.26 & -1.35 \\
Composite   & 50:50 & 120 & 1.05 & 5000 & 0.44 & 2.60 & 23.3  & 27.5 & -1.22 & -1.26 \\
Composite   & 50:50 & $T_{var}$$^{\footnotesize d}$& $v_{\rm var}$$^{\footnotesize e}$ & 2000 & 0.84 &3.89& 13.0& 21.3 & -1.07 & -1.51 \\
Composite   & 50:50 & $T_{var}$$^{\footnotesize d}$& $v_{\rm var}$$^{\footnotesize e}$ & 5000 & 0.72 & 3.48 & 14.3 & 21.9 & -1.03 & -1.40 \\
Composite   & 10:90 & $T_{var}$$^{\footnotesize d}$& $v_{\rm var}$$^{\footnotesize e}$ & 5000 & 0.37 & 1.76 & 25.4 & 37.7 & -1.33 & -1.53 \\
\hline
\end{tabular}
\end{table}
\end{landscape}

\noindent \underline{Footnotes to Table 2 :}

\noindent
 $^{\footnotesize a}$ {\small Relative contribution (in percent) of the
nuclear CO production rate $Q_{\rm CO }^N$ and extended CO
production rate $Q_{\rm CO }^E$ to the total CO production rate
$Q_{\rm CO }$ = $Q_{\rm CO }^N$ + $Q_{\rm CO }^E$ = 2.1
$\times$10$^{30}$ s$^{-1}$.}

\noindent $^{\footnotesize b}$ {\small Slope (i.e., power index)
of the variation of the visibility amplitude with $uv$-radius
between 20 and 150 m.}

\noindent $^{\footnotesize c}$ {\small Opacity effects are not
considered in radiative transfer calculations.}

\noindent $^{\footnotesize d}$ {\small Variable kinetic
temperature $T_{\rm var}$: $T = 20+50 \times \log(r[\rm km]/100)$
K for $100 < r < 10000 $ km, $T = 120$ K for $r > 10000$ km,
 $T = 20$ K for $r < 100$ km.}

\noindent
 $^{\footnotesize e}$ {\small Variable expansion velocity $v_{\rm
var}$: $v_{\rm exp} = 0.9+0.15 \times \log(r[\rm km]/1000)$ km
s$^{-1}$ for $r > 1000 $ km,  $v_{\rm exp}$ = 0.9 km s$^{-1}$ for
$r < 1000$ km.}

\noindent $^{\footnotesize f}$ {\small Variable kinetic
temperature $T_{\rm var(alt)}$: $T = 20+50 \times \log(r[\rm
km]/100)$ K for $1500 < r < 10000 $ km, $T = 120$ K for $r >
10000$ km, $T = 79$ K for $r < 1500$ km.}

\newpage

\begin{table}
\caption{CO IR line fluxes in March 1997.} \label{tab-intIR}
\begin{center}
\begin{tabular}{lllccccc}
\hline \noalign{\smallskip}
Date & Line & Comment & $Q_{\rm CO}$ &$\delta$Dec$^{\footnotesize a}$ &  $F_{inner}$$^{\footnotesize b, c}$ & $F_{outer}$$^{\footnotesize b, d}$ & $F_{outer}$/$F_{inner}$ \\
& & & (s$^{-1}$) & (\arcsec) & (W m$^{-2}$) & (W m$^{-2}$) & \\
\hline \noalign{\smallskip}
3.80 March UT & R2 & Measured$^{\footnotesize b}$ &  &  & 43 $\pm$ 1.5 & 46.8 $\pm$ 2.9 & 1.09 $\pm$ 0.08 \\
        &    & Panel 3/Iso$^{\footnotesize b}$    & 2$\times$10$^{30}$ & 0 & 43.1/45.5 & 37.8/36.6 & 0.88/0.80\\
        &    & Panel 3/Iso$^{\footnotesize b}$    & 2$\times$10$^{30}$ & --1.2 & 40.7/39.1 & 37.4/35.8 & 0.92/0.92 \\
        &    & Panel 3/Iso$^{\footnotesize b}$    & 2$\times$10$^{30}$ & --2.4 & 31.5/28.6 & 33.6/33.7 & 1.07/1.18 \\
        \hline
5.99 March UT & R6 & Measured$^{\footnotesize b}$ &  &  & 20.6 $\pm$ 1.6 & 21.4 $\pm$ 1.4 & 1.04  $\pm$ 0.11 \\
        &    & Panel 10/Iso$^{\footnotesize b}$    & 2$\times$10$^{30}$ & 0 & 36.9/34.9 & 31.2/26.1  & 0.85/0.75 \\
        &    & Panel 10/Iso$^{\footnotesize b}$   & 2$\times$10$^{30}$ & --1.2 & 26.8/29.2 & 29.8/25.6 & 1.11/0.88 \\
        &    & Panel 10/Iso$^{\footnotesize b}$    & 2$\times$10$^{30}$ & --2.4 & 20.9/20.6 &  26.7/24.1 & 1.28/1.17 \\
\hline
\end{tabular}
\end{center}

$^{\footnotesize a}$ Slit offset in declination. The slit is
oriented East-West.

$^{\footnotesize b}$ Measured intensities are from \citet{bro03}.
Modelled intensities were computed using the brightness
distributions of Panels 3 and 10 of Fig~\ref{plot-map}, for 3 and
5 March, respectively (first number in columns 6--8) and an
isotropic model (referred to as Iso, second number in columns
6--8). A seeing of 1.78\arcsec~ and 1.7\arcsec~ \citep{bro03}, for
3 and 5 March, respectively.

$^{\footnotesize c}$ Line intensity in a
2\arcsec$\times$1\arcsec~box in RA$\times$Dec centred on the peak
of CO brightness along the slit.

$^{\footnotesize d}$ Line flux in East-West averaged
5\arcsec$\times$1\arcsec~boxes beginning 3 \arcsec~from peak.

\end{table}

\clearpage

\clearpage

\clearpage

\clearpage

\clearpage

\clearpage

\clearpage

\clearpage

\clearpage

\clearpage

\end{document}